\RequirePackage{fix-cm}
\documentclass[twocolumn]{svjour3}          
\smartqed  

\usepackage{graphicx}
\usepackage{amsmath}
\usepackage{esint}
\usepackage{dirtytalk}
\usepackage[normalem]{ulem}
\usepackage{natbib}
\usepackage[caption=false,font=scriptsize]{subfig}
\usepackage{epstopdf}
\usepackage{multirow}
\usepackage[dvipsnames]{xcolor}
\usepackage[colorlinks=true,citecolor=blue,linkcolor=ForestGreen]{hyperref}
\usepackage{mwe}
\usepackage{makecell}
\usepackage{hhline}
\usepackage{float}
\usepackage{bookmark}
\usepackage{tabu}
\usepackage{ulem} 
\usepackage{amsfonts} 
\usepackage{physics} 
\usepackage{relsize} 
\usepackage[bottom]{footmisc} 
\usepackage{enumitem} 
\setlist[itemize]{parsep=0pt, topsep=0pt, itemsep=0pt} 
\usepackage{booktabs} 
\usepackage{threeparttable} 
\usepackage{multirow} 
\usepackage{pdflscape} 
\usepackage[counterclockwise]{rotating} 
\usepackage{cuted} 
\usepackage{upgreek} 

\begin{document}
	
\title{\textbf{ \normalsize The official version of this paper can be downloaded at \href{https://link.springer.com/article/10.1007/s00158-022-03269-y} {https://doi.org/10.1007/s00158-022-03269-y}.} \vspace{2em} \\
Direction-Oriented Stress-Constrained Topology Optimization of Orthotropic Materials}
\author{Ahmed Moter \and Mohamed Abdelhamid \and Aleksander Czekanski}

\institute{A. Moter \at
	Department of Mechanical Engineering, Lassonde School of Engineering, York University, Toronto, Ontario, Canada \\
	\and
	M. Abdelhamid \at
	Department of Mechanical Engineering, Lassonde School of Engineering, York University, Toronto, Ontario, Canada \\
	\and
	A. Czekanski \at
	Department of Mechanical Engineering, Lassonde School of Engineering, York University, Toronto, Ontario, Canada \\
	\email{alex.czekanski@lassonde.yorku.ca}
}

\date{Received: date / Accepted: date}

\maketitle

\interfootnotelinepenalty=10000

\begin{abstract}
Efficient optimization of topology and raster angle has shown unprecedented enhancements in the mechanical properties of 3D printed materials. Topology optimization helps reduce the waste of raw material in the fabrication of 3D printed parts, thus decreasing production costs associated with manufacturing lighter structures. Fiber orientation plays an important role in increasing the stiffness of a structure. This paper develops and tests a new method for handling stress constraints in topology and fiber orientation optimization of 3D printed orthotropic structures. The stress constraints are coupled with an objective function that maximizes stiffness. This is accomplished by using the modified solid isotropic material with penalization method with the method of moving asymptotes as the mathematical optimizer. Each element has a fictitious density and an angle as the main design variables. To reduce the number of stress constraints and thus the computational cost, a new clustering strategy is employed in which the highest stresses in the principal material coordinates are grouped separately into two clusters using an adjusted $P$-norm. A detailed description of the formulation and sensitivity analysis is discussed. While we present an analysis of 2D structures in the numerical examples section, the method can also be used for 3D structures, as the formulation is generic. Our results show that this method can produce efficient structures suitable for 3D printing while thresholding the stresses.
\end{abstract}

\section{Introduction}
\label{sec:intro}
\subsection{Additive Manufacturing}
Technological advancements usually increase the stringency and sophistication of the required specifications of a fabricated structure \citep{anand}. A fundamental engineering advancement that completely changed the targets set by manufacturers is \textbf{a}dditive \textbf{m}anufacturing (AM). During the AM process, a 3D CAD model is converted to a real-world commercial product. This process occurs without any human interference, as the 3D printer accurately adheres the material layer by layer until the final product is produced \citep{ngo}.

Before manufacturing any product, the structure--process--property relationship must first be defined. This relationship is decisive when anisotropy is expected to show up in the part during the manufacturing process as a result of void formation, build orientation, patterning, and interfacing techniques, all of which have  made this area of research crucial in the development of 3D printing processes \citep{barclift2012examining, ajoku2006investigating}. Some printing methods produce anisotropic properties in the manufactured structures such as fused deposition modeling, polyjet direct 3D printing, and laser sintering \citep{ajoku2006investigating}. Methods used to produce metal components - including selective laser melting and stereolithography - show insignificant anisotropy within the printing layers. The anisotropy introduced into the part using these manufacturing methods is usually in the direction perpendicular to the print bed \citep{dulieu, kempen2012mechanical}.

The aforementioned 3D printing methods would be more efficient if the parts being produced were structurally optimized before beginning the printing process, as industry tends to favor cheap, lightweight, and easily manufactured engineering structures. Structural optimization has three main types: shape optimization, size optimization, and \textbf{t}opology \textbf{o}ptimization (TO). TO has proven its efficiency over the former methods in that it offers the freedom of choosing the optimal material distribution in a defined design space without constraining either the dimension or the shape of the structure \citep{christensen_2008}. TO is a mathematical design tool requiring post-processing and analysis. The aim is to facilitate further design work from a stronger starting point. Thus, TO has been universally adopted as the main methodology for the design of AM parts as it can produce complex and efficient structures that can easily be 3D printed \citep{gao2015status, liu2017deposition}. 

\subsection{Topology Optimization Approaches}
TO was first developed in 1988 by \citet{bendsoe1988generating} using the homogenization method. Later, its popularity declined significantly owing to the development of density-based approaches. The latter revolutionized the field of TO due to its ease of implementation, low computational complexity, low failure rate, and low number of predefined parameters \citep{cui2016level}. The \textbf{s}olid \textbf{i}sotropic \textbf{m}aterial with \textbf{p}enalization (SIMP) method, the most well-known of all density-based approaches, was developed in 1989 by \citet{bendsoe1989optimal}. Evolutionary optimization methods are well-known for handling problems with large numbers of design variables, and therefore were tried in a TO context by \citet{xie1993simple}. Nonetheless, it is not the best approach because it has been shown to be non-convergent in some problems \citep{maute2013topology}. Moreover, although it does perform a global search, it is not guaranteed to find the global optimum. A relatively recent development in TO is the boundary variation methods \citep{sethian2000structural}. The drawbacks of level-set methods are that the geometry cannot evolve outside the existing boundaries (i.e. no new holes can be generated at points surrounded by solid material), and therefore they have to be accompanied by some technique to overcome this issue.

Despite the diversity of the aforementioned methods, there is still a lot of unanswered questions when it comes to multi-scale problems. Recent works include those of \citet{gao_luo_xia_gao_2019}, who once again called upon the homogenization approach, since it is capable of handling multi-scale optimization problems. \citet{gao_luo_xia_gao_2019} developed a concurrent method that links macro- and micro-scale optimization problems. This method dealt with multi-scale composites, however, it is not logical to use in determining fiber angle orientation because micro-structures have the same topology at the end of the solution and do not show any indication of the fiber angle orientation. The design variables of this method are the densities of the micro- and macro-structures, however, the solution algorithm is so complex that it will be very difficult to add the fiber angle as a new design variable. \citet{papapetrou2020stiffness} introduced a three-stage sequential optimizing process, in which isotropic design, continuous fiber orientation, and functionally graded discrete orientation design, respectively, are provided for each phase. \citet{peeters2015combining} used the lamination parameters method. This approach addressed multi-layered composites, not 3D printed polymers. According to \citet{peeters2015combining}, the method requires major simplifications and assumptions to be able to reach a feasible result, which - as the authors mention - is not manufacturable. \citet{jia2008topology} introduced the modified SIMP method in 2008. This method builds on the work of \citet{bendsoe1989optimal} published in 1989 in its use of the fictitious density method. It is privileged with being easy to implement and its ability to produce manufacturable products. This method was thereafter used by \citet{jiang2019continuous} and \citet{ramsey2019topology}.

As for the mathematical solvers, the optimality criteria method was the first method used to solve the topology optimization problem due to its simplicity and ease of programming \citep{bendsoe1989optimal}. \citet{Yang1997} generalized the topology optimization formulation to be solved using the mathematical programming method. \citet{jia2008topology} and \citet{jiang2019continuous} used {\tt fmincon} and the modified feasible direction methods, respectively, in their search for the minima. Both methods are locally convergent and can be stuck at a certain point where the function is not decreasing anymore and produces non-optimal results. \citet{ramsey2019topology} used the \textbf{m}ethod of \textbf{m}oving \textbf{a}symptotes (MMA) method \citep{svanberg2007mma}. MMA is considered more efficient and requires fewer iterations in solving the compliance minimization problem \citep{bendsoe2004topology}. Currently available commercial software has yet to develop modules that can deal with multi-scale problems \citep{jiang2019continuous}. Most of the TO modules implemented in commercial software use SIMP or the level-set methods, which allow solving TO problems for 3D printed polymers. However, it either optimizes the structural topology for a given fiber orientation or the fiber orientation for a given topology. In other words, coupling TO and fiber orientation is beyond commercial software capabilities.

\subsection{Stress Constraints in Topology Optimization}
Many research articles have been published about coupling topology and raster angle optimization \citep{jiang2019continuous, ramsey2019topology, liu2021stress}. However, all this work was focused on discussing global response constraints, namely compliance and volume. Stress constraints have not yet been addressed, as they require additional steps to overcome three challenges; the singularity phenomenon \citep{kirsch1990singular, cheng1992study, rozvany1994singular, han2021topology}, the local nature of stresses which results in a huge number of stress constraints, and the high non-linearity of stresses \citep{lin2009adaptive, holmberg2013stress, liu2021stress, Abdelhamid2021}. Only the first two problems will be directly addressed in this paper with some commentary on the third.

The stress constraints of an orthotropic material are typically higher than that of isotropic materials, as the stresses in different directions are independent and should be controlled. The singularity phenomenon occurs when the globally optimal solution is a degenerate subspace of dimension less than that of the original feasible space of the problem. For instance, when solving a truss problem, some members' areas go to zero, while these members are essential and cannot be removed. Similarly, when dealing with a continuum, an element with a low density can still have a strain and thus generates a high stress value in this area while it should be zero. This can be fixed by using the $\varepsilon$-relaxation techniques mentioned in \cite{cheng1997varepsilon} or by using the stress penalization approach proposed by \cite{bruggi2008alternative}. We used the latter method, as it helps solve the singularity problem and penalizes the stress values to avoid its non-linear behavior (see subsection \ref{stressp}). 

The local nature of stresses results in the number of constraints being equal to multiples of the number of elements depending on the failure criteria used. For instance, if the failure criteria is based on stresses along and perpendicular to the fibers only, the number of constraints will be double the elements number. If shear stress is also considered, the constraints will be triple the number of elements. This drastically compromises the optimization process, as it becomes extremely computationally expensive. Solutions proposed for this problem include the global stress measure introduced by \citet{duysinx1998new} and the block aggregation method introduced by \citet{paris2010block}. The former method reduces all the stress constraints in one direction to only one constraint, which is computationally efficient but not accurate and does not always control the stress \citep{xu2021stress}. The latter method is the basis of the clustering technique introduced by \citet{holmberg2013stress}. The maximum stress cluster technique used in this work is based on the latter method together with the minimax approach \citep{brittain2012minmax}. In brief, the maximum stresses in the first principal direction will be integrated into a cluster. The number of stress evaluation points is chosen based on the number of finite elements. A good ratio based on numerical experiments was found to be $2.5\%$ of the elements in the constraint cluster. The stress values of these elements are integrated using the $P$-norm global stress measure. Stresses in the second principal direction are treated in the same way. 

This work focuses on a new technique called the \textit{maximum stress level clustering technique}, which adds stress constraints to the problem of simultaneously optimizing topology as well as fiber orientation to avoid high stresses in orthotropic material. Stress constraints will be based on the maximum principal stress failure theory to control stress values in the directions of the principal material coordinates.

\section{Methods}
\subsection{Stress Problem Formulation}

We extend the conventional TO problem of compliance minimization subjected to a volume constraint with element densities as design variables to a problem that has the density $\rho_{e}$ and fiber angle $\theta_{e}$ of each finite element as design variables and stress constraints along the fibers $\sigma_1$ and perpendicular to the fibers $\sigma_2$. All the design variables of the whole domain are collected in the vector \(\mathbf{x}\).
\begin{equation}
	\begin{array}{c}
		\vspace{5px}
		\text { minimize : } c(\boldsymbol{\uprho}, \boldsymbol{\uptheta})=\mathbf{U}(\boldsymbol{\uprho}, \boldsymbol{\uptheta})^T \, \mathbf{F}, \\
		\text { subject to : }
		\left\{
		\begin{array}{c}
			\vspace{5px}
			\sum_{j=1}^{N}V_j/V_0 \leq V_f, \\
			\mathbf{K}(\boldsymbol{\uprho}, \boldsymbol{\uptheta}) \, \mathbf{U}(\boldsymbol{\uprho},	\vspace{5px}\boldsymbol{\uptheta})=\mathbf{F}, \\
			\vspace{5px}
			\sigma_i^{PN}(\mathbf{x}) \leq \text{min} (\sigma^C_i, \sigma^T_i), \\
			\vspace{5px}
			0 \leq \rho_{\mathrm{e}} \leq 1, \\
			\vspace{5px}
			-\pi \leq \theta_{\mathrm{e}} \leq  \pi. \vspace{-5px}
		\end{array}
	\right.
	\end{array}
 \label{eq:general_eqtn}
\end{equation}

\noindent where \(c\) is the compliance of the structure, \(\mathbf{U}\) is the global displacement vector which is a function of the density and fiber angle design variables $\boldsymbol{\uprho}$ and $\boldsymbol{\uptheta}$, \(\mathbf{K}\) is the global stiffness matrix, \(N\) is the total number of finite elements, \(V_j\) is the volume of element \(j\), \( V_0 \) is the volume of the original design domain, \( V_f \) is the prescribed volume fraction, \(\mathbf{F}\) is the global force vector, and \(\rho_e\) and $\theta_e$ are the density and fiber angle of element \(e\), respectively. The vectors $\boldsymbol{\uprho} = [\rho_{1},\rho_{2},..,\rho_{N}]$ and $\boldsymbol{\uptheta} = [\theta_{1},\theta_{2},..,\theta_{N}]$ are defined by element density design variables $\rho_{e}, \, e = 1,...,N$ and element material angle design variables $\theta_{e}, \, e = 1,...,N$ respectively. ${\sigma^{PN}_i(\mathbf{x})}$ is the modified $P$-norm stress measure\footnote{The modified $P$-norm described later in Eq. \ref{eq:pnorm_general} might produce complex stress values if applied to negative stresses with odd $P$ values, so to avoid this issue we always use even $P$ values.} for the cluster of stresses in the principal material coordinate $i$ (discussed in detail in subsection \ref{ssec:stress_calc}). Lastly $\sigma^C_i$ and $\sigma^T_i$ are the ultimate compressive and tensile stresses in the principal material direction $i$, respectively. Note that in Eq. \ref{eq:general_eqtn}, the governing equation components are defined in the global coordinate system ($x$, $y$, and $z$) while the stress components are defined in the principal material coordinate system ($1$, $2$, and $3$).

\subsection{Penalization}
\subsubsection{Stiffness Penalization}
A penalization function is used to make intermediate density elements disproportionately expensive in order to create black-and-white structures. In this paper, the modified SIMP is used to penalize stiffness for intermediate density elements \citep{jia2008topology}. The global stiffness matrix $\mathbf{K} = \mathbf{K}(\boldsymbol{\uprho}, \boldsymbol{\uptheta})$ is mapped from each element's stiffness matrix as follows: \begin{equation}
	\mathbf{K}(\boldsymbol{\uprho}, \boldsymbol{\uptheta}) = \sum_{e=1}^N \eta_K (\rho_{e}) \, \mathbf{k}_e(\theta_e).
\end{equation}

\noindent where $\eta_{K}(\rho_{e})=\rho_{e}^p$ and $p$ is the penalization parameter. The fiber direction angle is added as a design variable and according to classical laminate theory for plane stress composites \citep[p.~80]{herakovich1998mechanics}:
\begingroup
\allowdisplaybreaks
\begin{gather}
	\boldsymbol{\hat{\upsigma}}^e = \mathbf{E}^\mathbf{'}_e \, \boldsymbol{\hat{\upepsilon}}^e, \\
	\mathbf{E}^\mathbf{'}_e(\theta_e) = \mathbf{T}_{\mathbf{1}e}(\theta_e)^{-1} \, \mathbf{E} \, \mathbf{T}_{\mathbf{2}e}(\theta_e), \\
	\mathbf{T}_{\mathbf{1}e} = \begin{bmatrix}
		c^2 & s^2 & 2cs \\
		s^2 & c^2 & -2cs \\
		-cs& cs & c^2-s^2\\
	\end{bmatrix}, \\
	\mathbf{E} = \begin{bmatrix}
	\dfrac{E_1} {1-\nu_{12} \nu_{21}} & \dfrac{\nu_{12} E_2}{1-\nu_{12} \nu_{21}}  & 0 \\[8pt]
	\dfrac{\nu_{21} E_1}{1-\nu_{12} \nu_{21}} & \dfrac{E_2}{1-\nu_{12} \nu_{21}} & 0 \\[8pt]
	0 & 0 & G_{12} \\
	\end{bmatrix}, \\
	\mathbf{T}_{\mathbf{2}e} = \begin{bmatrix}
		c^2 & s^2 & cs \\
		s^2 & c^2 & -cs \\
		-2cs & 2cs & c^2-s^2 \\
	\end{bmatrix}, \\
	\mathbf{k}_e = \mathbf{B}_e^T \, \mathbf{E}^\mathbf{'}_e(\theta_e) \, \mathbf{B}_e = \mathbf{B}_e^T \, \mathbf{T}_{\mathbf{1}e}(\theta_e)^{-1} \, \mathbf{E} \, \mathbf{T}_{\mathbf{2}e}(\theta_e) \, \mathbf{B}_e.
\end{gather}
\endgroup

\noindent where $\boldsymbol{\hat{\upsigma}}$ and $\boldsymbol{\hat{\upepsilon}}$ are the stress and strain vectors in the global coordinates, $\mathbf{E}^\mathbf{'}_e$ is the transformed constitutive matrix (i.e. in the global coordinates), $\mathbf{T}_{\mathbf{1}e}$ and $\mathbf{T}_{\mathbf{2}e}$ are the stress and strain transformation matrices, $\mathbf{B}_e$ is the strain-displacement matrix, \(\mathbf{E}\) is the original constitutive matrix (i.e. in the principal material coordinates), and lastly $c$ and $s$ represent \(cos(\theta_e) \textrm{ and } sin(\theta_e)\).

\subsubsection{Stress Penalization} \label{stressp}

The stress tensor of a 2D finite element is defined as follows \citep{bathe2006finite}: 
\begin{equation}
    \begin{split}
        \boldsymbol{\hat{\upsigma}}^e(\mathbf{x}) & = \begin{bmatrix} \hat{\sigma}^e_{xx} \\ \hat{\sigma}^e_{yy} \\ \hat{\tau}^e_{xy} \end{bmatrix} = \mathbf{E}^\mathbf{'}_e(\theta_e) \, \mathbf{B}_e \, \mathbf{u}_e(\rho_e, \theta_e) = \\
        & = \mathbf{T}_{\mathbf{1}e}(\theta_e)^{-1} \, \mathbf{E} \, \mathbf{T}_{\mathbf{2}e}(\theta_e) \, \mathbf{B}_e \, \mathbf{u}_e(\rho_e, \theta_e).
    \end{split}
\end{equation}

\noindent where $\mathbf{u}_e(\rho_e, \theta_e)$ is the displacement vector of finite element $e$ in the global coordinates. Stresses are penalized as well to obtain intermediate stress values that depend on the fictitious density of each element. The stress penalization functions in literature are diverse; we chose the method suggested by \citet{bruggi2008alternative} as it proved its efficiency when coupled with the $P$-norm stress measure in \cite{holmberg2013stress}. Stress penalization is achieved as follows: 
\begin{equation}
    \begin{split}
       \boldsymbol{\upsigma}^e(\mathbf{x}) & = \eta_S(\rho_e) \, \mathbf{T}_{\mathbf{1}e}(\theta_e) \, \boldsymbol{\hat{\upsigma}}^e(\mathbf{x}) = \\
       & = \eta_S(\rho_e) \, \mathbf{E} \, \mathbf{T}_{\mathbf{2}e}(\theta_e) \, \mathbf{B}_e \, \mathbf{u}_e(\rho_e, \theta_e).
    \end{split}
	\label{eq:penalizedstress}
\end{equation}

\noindent where $\eta_{S}(\rho_e) = {\rho_e}^{\tfrac{1}{2}}$. Note that due to the existence of $\mathbf{T}_{\mathbf{1}e}$ in Eq. \ref{eq:penalizedstress}, $\boldsymbol{\upsigma}^e(\mathbf{x})$ is now defined in the principal material coordinates. This stress penalization gives an artificially lower stress for intermediate densities while removing the penalization effect for $\rho_e = 1$. Stress singularities are eliminated due to the following fact \citep{holmberg2013stress}:
\begin{equation}
	\lim _{\rho_e \rightarrow 0} \boldsymbol{\upsigma}^e(\mathbf{x}) = \mathbf{0}.
\end{equation}

It's worth noting that, in this work, a superscript appended to stresses indicates the corresponding finite element/evaluation point while a subscript indicates a specific component of the stress vector with respect to the coordinate system. To indicate raising the stress to a certain power, we will use parentheses around the stress value as in $(\sigma_\square^\square)^2$ to indicate the stress squared for example. The same applies to strain notation.

\subsection{Stress Calculation}
\label{ssec:stress_calc}

\subsubsection{Failure Criteria}
Composite failure methods are varied at the fiber/matrix level. Fiber fracture, fiber buckling, fiber splitting, matrix cracking, and radial cracks are all failure mechanisms at the fiber level. Transverse cracks in planes parallel to the fibers or perpendicular to the fibers in the form of delamination between the laminate layers are examples of laminate failures \citep{herakovich1998mechanics, chen2016modelling, mouritz2012introduction}.

The most serious form of fiber fracture is transverse fiber fracture, which is represented by the first principal material coordinate in the current work, as fibers are typically the primary load carriers. The second principal material coordinate represents the perpendicular or delamination direction. The maximum tensile and compressive stresses of fibers are usually determined by the fiber ultimate strength in each direction. Failure expectancy can be determined through multiple theories, namely maximum stress, maximum strain, Tsai--Hill, and tensor polynomial failure criteria. We use the maximum stress failure criteria in this work because it only requires the stress values in the two principal material coordinates. Shear values are ignored in this study.

Maximum stress failure theory requires all individual stress components to be less than the ultimate stress value in the assumed direction. Thus, the failure criteria can be written as follows:
\begin{equation}
	\begin{aligned}
		\sigma_i^C < \sigma^e_i < \sigma_i^T, \\
		\tau^e_{12}<Q.
	\end{aligned}
\end{equation}

\noindent where $\sigma_i^T$ and $\sigma_i^C$ are the maximum allowed tensile and compressive stresses in the principal material coordinate $i$. $\sigma^e_i$ is the stress at finite element $e$ in the principal material coordinate $j$. $\tau_{12}$ and $Q$ are the shear stress and the maximum allowed shear stress, but - as mentioned before - this constraint is out of the scope of this study.

To decrease the number of stress constraints, the absolute value of the stresses will be used, hence a single ultimate stress will be assumed for both tension and compression. Hence, the stress constraints are as follows: 
\begin{equation}
	\begin{aligned}
		|\sigma^e_i| < \min (\sigma_i^C, \sigma_i^T).
	\end{aligned}
\end{equation}

\subsubsection{Stress Measure} \label{ssec:strss_msre}
As mentioned in the introduction, the local nature of stresses prompts us to use a global stress technique together with stress clustering techniques in order to control the stresses using a low number of constraints and still obtain accurate results \citep{holmberg2013stress}. Stresses can be constrained using three techniques: local, global, or clusters. A local stress measure means that every element will be accompanied by two stress constraints (while ignoring shear), which is not computationally efficient. Global constraining means only one constraint will be added to the whole mesh using a global measure such as $P$-norm \citep{duysinx1998new}. The global $P$-norm stress measure in the principal material coordinate $i$ takes the following form: 
\begin{equation}
	\sigma^{PN}_i(\mathbf{x})=\left( \frac{1}{N} \sum_{a ={1}}^N \bigg( \sigma_i^a(\mathbf{x}) \bigg)^P \right)^{\tfrac{1}{P}}.
	\label{eq:pnorm_general}
\end{equation}

\noindent where $P$ is the $P$-norm coefficient and $a$ designates the finite element or the evaluation point. The formulation of Eq. \ref{eq:pnorm_general} forces $\sigma^{PN}_i$ to approach the maximum stress value in the whole domain only when $P \rightarrow \infty$ \citep{duysinx1998new}.
\begin{equation}
	\lim _{P \rightarrow \infty}\left( \frac{1}{N} \sum_{a = 1}^N \bigg( \sigma_i^a(\mathbf{x}) \bigg)^P \right)^{\tfrac{1}{P}} = \max_a \big( \sigma^a_i(\mathbf{x}) \big).
	\label{eq:pnorm_limit}
\end{equation}

From Eq. \ref{eq:pnorm_limit} we can conclude that the global stress measure will be effective when used with a low count of stress evaluation points but will not be accurate if used with the whole mesh. This means that such technique would work well when the stresses are divided into clusters. Each cluster will have its own $P$-norm measure, an approach that yields a better approximation of the highest stress value in that cluster. However, it would still underestimate the stress values in the cluster depending on the amount of stress evaluation points that dominate the cluster. Nevertheless, the $P$-norm value will always underestimate the maximum stress in the cluster.

Clusters can be arranged by two methods. The first one uses \textit{the stress level technique}, whereby stresses are arranged in descending order and then added to clusters from the highest to the lowest value. Each cluster will have stress evaluation points equal to the total number of elements $N$ divided by the number of constraints $N_c$ (i.e. number of clusters). It can be written mathematically in the form of Eq. \ref{eq:descending} \citep{holmberg2013stress}. The second technique is \textit{the distributed stress method}, where the elements are arranged in descending order and then each cluster gets an element, i.e. the first element is inserted into the first cluster, the second element is inserted into the second cluster, and so on, until each cluster has one point \citep{holmberg2013stress}. Then, the loop starts again until all the evaluation points are placed in their corresponding clusters. This technique helps with averaging the global stress measure in each cluster. 

In this work, we adopt a new technique combining the minmax technique used in \citet{brittain2012minmax} and the stress level technique. For each principal material coordinate, the first cluster which has the maximum stresses will be the only constraint in that coordinate, taking into account that the number of stress evaluation points per cluster $N_s$ should be at least 2.5\% of the total number of elements. This is henceforth referred to as the \textit{maximum stress level clustering technique}. 

\begin{strip}
	\begin{gather}
		\underbrace{\sigma_{1}^1 \geq \sigma_{1}^2 \geq \sigma_1^{3} \geq \ldots . . \geq \sigma_1^{\tfrac{N}{N_c}}}_{\mathrm{Cluster \, 1}} \geq \underbrace{\cdots \cdots \geq \sigma_1^{\tfrac{2N}{N_c}}}_{\mathrm{Cluster \, 2}} \geq \underbrace{\cdots \cdots
		\geq \sigma_1^{\tfrac{(N_c-1)N}{N_c}}}_{\mathrm{Cluster \,} N_c-1} \geq \underbrace{\cdots \geq \sigma_1^N}_{N_c}.
		\label{eq:descending}
	\end{gather}
\end{strip}


\subsection{Optimization Process}
Since compliance minimization is considered a non-linear programming problem, it can be solved using any of the well-known optimization methods, such as Newton or quasi-Newton. However, owing to the large number of design variables being considered here, MMA has already proven to be well suited for solving such problems \citep{svanberg2007mma}. The main idea of this method is instead of solving the original non-convex problem, a set of approximated strictly convex subproblems are solved sequentially. These strictly convex functions are chosen based on moving lower and upper asymptotes. Moreover, it can be globally convergent (i.e. GCMMA) and guarantees convergence as long as a feasible solution exists. 

\subsection{Objective Function, Constraints, and Sensitivity Analysis}
As the MMA is a gradient-based method, gradients of the objective function and constraints are needed before the optimization process starts. The compliance and its derivatives w.r.t. the design variables of finite element $e$ are calculated according to \cite{jiang2019continuous} as follows:
\begin{gather}
	c(\boldsymbol{\uprho}, \boldsymbol{\uptheta}) = \sum_{e = 1}^{N} \rho_e^p \, \mathbf{u}_e^T \, \mathbf{k}_e \, \mathbf{u}_e, \\
	\frac{\partial c}{\partial \rho_{e}} = - \, p \, \rho_e^{p-1} \, \mathbf{u}_e^T \, \mathbf{k}_e \, \mathbf{u}_e, \\
	\begin{split}
	    	\frac{\partial c}{\partial \theta_{e}} = - \, \rho_e^p \, \mathbf{u}_e^T \mathbf{B}^T & \left( \pdv{ \mathbf{T}_{\mathbf{1}e}^{-1}}{\theta_e} \, \mathbf{E} \, \mathbf{T}_{\mathbf{2}e} \right. \\
	    & \left. + \, \mathbf{T}_{\mathbf{1}e}^{-1} \, \mathbf{E} \, \pdv{\mathbf{T}_{\mathbf{2}e}}{\theta_e}\right) \mathbf{B} \, \mathbf{u}_e.
	\end{split}
\end{gather}

To decrease the number of constraints, only one stress constraint will be added for stresses in both compression and tension. Hence, the absolute value of stress at each evaluation point must be calculated. To simplify the derivatives, the square root of the stress squared is taken instead of the absolute value as in Eq. \ref{eq:stressabsolute}. 
\begin{equation}
	\bar{\sigma}_i^a(\mathbf{x})= \sqrt{ \bigg( \sigma_i^a(\mathbf{x}) \bigg)^2 }.
	\label{eq:stressabsolute}
\end{equation}

\noindent where $\bar{\sigma}_i^a(\mathbf{x})$ is the absolute stress value in the principal material coordinate $i$ at stress evaluation point $a$. Using the stress level technique discussed in section \ref{ssec:strss_msre} and detailed in Eq. \ref{eq:pnorm_general}, the maximum stress cluster $P$-norm is defined as:
\begin{equation} \label{eq:clustering}
	\sigma_i^{PN}(\mathbf{x}) = \left( \frac{1}{N_s} \mathlarger{\sum}_{a \in \Omega_i} \bigg( \bar{\sigma}_i^a(\mathbf{x}) \bigg)^P \right)^{\tfrac{1}{P}}.
\end{equation} 

\noindent where $\Omega_i$ is the cluster domain that holds the maximum $N_s$ stress values in the principal material coordinate $i$. 

To calculate the derivative of the clustered $P$-norm stress measure defined in Eq. \ref{eq:clustering} w.r.t. the design variables, the chain rule is used as follows:
\begin{equation}
	\frac{\partial \sigma_i^{P N}}{\partial x_e} = \sum_{a \in \Omega_i} \frac{\partial \sigma_i^{P N}}{\partial \bar{\sigma}_i^a} \, \frac{\partial \bar{\sigma}_i^a}{\partial \sigma_i^a} \, \frac{\partial \sigma_i^a}{\partial x_e}.
	\label{eq:pnorm_derivatives}
\end{equation}

\noindent where $x_e$ can be $\rho_e$ or $\theta_e$. The right hand side of Eq. \ref{eq:pnorm_derivatives} consists of three multiplied derivatives, the first is a derivative of Eq. \ref{eq:clustering} w.r.t. the absolute stress:
\begin{equation}
	\begin{aligned}
		\frac{\partial \sigma_i^{PN}}{\partial \bar{\sigma}_i^a} = \left( \frac{1}{N_s} \sum_{a \in \Omega_i} \big( \bar{\sigma}_i^a \big)^P \right)^{\left(\tfrac{1}{P}-1\right)} \cdot \frac{1}{N_s} \big( \bar{\sigma}_i^a \big)^{P-1}.
	\end{aligned}
\end{equation}

The second derivative can be obtained from Eq. \ref{eq:stressabsolute} as follows:
\begin{equation}
    \frac{\partial \bar{\sigma}_i^a}{\partial \sigma_i^a} = \frac{\sigma_i^a}{\sqrt{\big( \sigma_i^a \big)^2}}.
\end{equation}

Before calculating the third derivative, we will perform some algebraic manipulations to separate the vector components of $\boldsymbol{\upsigma}^a$ into $\sigma_i^a$ where $i$ indicates the principal material coordinate as discussed earlier. Starting from Eq. \ref{eq:penalizedstress}:
\begin{equation}
    \sigma_i^a = \, \eta_S(\rho_e) \, \mathbf{E}^i \, \mathbf{T}_{\mathbf{2}e}(\theta_e) \, \mathbf{B}_e \, \mathbf{u}_e(\rho_e, \theta_e).
\end{equation}

\noindent such that $\mathbf{E}^i$ indicates row $i$ of $\mathbf{E}$. Hence, the third derivative in the right hand side of Eq. \ref{eq:pnorm_derivatives} is:
\begin{gather}
    \frac{\partial \sigma_i^a}{\partial x_e} = \pdv{ \big( \eta_S \, \mathbf{E}^i \, \mathbf{T}_{\mathbf{2}a} \, \mathbf{B}_a \big) }{x_e} \, \mathbf{u}_a + \eta_S \, \mathbf{E}^i \, \mathbf{T}_{\mathbf{2}a} \, \mathbf{B}_a \, \frac{\partial \mathbf{u}_a}{\partial x_e}.
    \label{eq:sigma_a_deriv}
\end{gather}

The derivative of displacements can be calculated from the global state equation:
\begin{gather}
    \frac{\partial \mathbf{U}}{\partial x_e} = \, - \, \mathbf{K}^{-1} \, \pdv{\mathbf{K}}{x_e} \, \mathbf{U}.
    \label{eq:disp_deriv}
\end{gather}

Substituting Eq. \ref{eq:sigma_a_deriv} into Eq. \ref{eq:pnorm_derivatives} as follows:
\begin{gather}
\begin{split}
\pdv{\sigma_i^{P N}}{x_e} = 
\sum_{a \in \Omega_i} \frac{\partial \sigma_i^{P N}}{\partial \bar{\sigma}_i^a} \, \frac{\partial \bar{\sigma}_i^a}{\partial \sigma_i^a}  & \bigg(\frac{\partial \big(\eta_S \, \mathbf{E}^i \, \mathbf{T}_{\mathbf{2} a} \, \mathbf{B}_a \big)}{\partial x_{e}} \, \mathbf{u}_a \\
& + \eta_S \, \mathbf{E}^i \, \mathbf{T}_{\mathbf{2} a} \, \mathbf{B}_a \pdv{\mathbf{u}_{a}}{x_e} \bigg).
\end{split}
\end{gather}

In topology optimization problems, the adjoint method is typically preferred over the direct method for calculating the sensitivities. This is mainly due to the large number of design variables compared to the relatively lower number of design constraints \citep{Tortorelli1994}. A linear system is typically solved to avoid inverting $\mathbf{K}$ where its decomposition, generated when solving $\mathbf{K} \mathbf{U} = \mathbf{F}$, can be reused \citep{nocedal2006numerical, kambampati2021discrete}. The adjoint equation is mathematically defined as:
\begin{gather}
\begin{split}
	\mathbf{K}(\boldsymbol{\uprho}, \boldsymbol{\uptheta}) \, \boldsymbol{\uplambda} =  \sum_{a \in \Omega_{i}} \frac{\partial \sigma_i^{P N}}{\partial \bar{\sigma}_i^a} \, \frac{\partial \bar{\sigma}_i^a}{\partial \sigma_i^a} \, \mathbf{B}_{a}^{T} \, \mathbf{T}_{2 a}^{T} \, (\mathbf{E}^{i})^{T}.
	\label{eq:adjoint}
\end{split}
\end{gather}

The linear system in Eq. \ref{eq:adjoint} is solved for the adjoint vector $\boldsymbol{\lambda}$. Eventually, Eq. \ref{eq:pnorm_derivatives} is transformed to Eqs. \ref{eq:sigmaderv1} and \ref{eq:sigmaderv2} as follows:

\begin{gather}
\begin{split}
	\frac{\partial \sigma_{i}^{P N}}{\partial \rho_{e}} & =  \sum_{a \in \Omega_{i}} \frac{\partial \sigma_i^{P N}}{\partial \bar{\sigma}_i^a} \, \frac{\partial \bar{\sigma}_i^a}{\partial \sigma_i^a} \frac{\partial \eta_S}{\partial \rho_{e}} \, \mathbf{E}^{i} \, \mathbf{T}_{\mathbf{2} a} \, \mathbf{B}_{a} \, \mathbf{u}_a \\
	& - \sum_{a \in \Omega_{i}} \frac{\partial \sigma_i^{P N}}{\partial \bar{\sigma}_i^a} \, \frac{\partial \bar{\sigma}_i^a}{\partial \sigma_i^a} \, \eta_{S} \, \boldsymbol{\uplambda}^T \,  \frac{\partial \mathbf{K}}{\partial \rho_{e}} \, \mathbf{u}_a,
	\label{eq:sigmaderv1}
\end{split} \\
\begin{split}
	\frac{\partial \sigma_{i}^{P N}}{\partial \theta_{e}} & = \sum_{a \in \Omega_{i}}  \frac{\partial \sigma_i^{P N}}{\partial \bar{\sigma}_i^a} \, \frac{\partial \bar{\sigma}_i^a}{\partial \sigma_i^a} \, \eta_S \, \mathbf{E}^i \, \frac{\partial \mathbf{T}_{\mathbf{2} a}}{\partial \theta_{e}} \, \mathbf{B}_a \, \mathbf{u}_a  \\
	& - \sum_{a \in \Omega_{i}}  \frac{\partial \sigma_i^{P N}}{\partial \bar{\sigma}_i^a} \, \frac{\partial \bar{\sigma}_i^a}{\partial \sigma_i^a} \, \eta_S \, \boldsymbol{\uplambda}^T
	\, \frac{\partial \mathbf{K}}{\partial \theta_{e}} \, \mathbf{u}_a.
	\label{eq:sigmaderv2}
\end{split}
\end{gather}

\subsection{Filtering}
Filtering is conventionally performed using sensitivity or density filtering methods \citep{bendsoe2004topology}. Filtering is crucial to solving two main problems that usually appear when using TO. The first is the mesh dependency problem, meaning that results are overly dependent on the mesh size \citep{sigmund1998numerical}. The second is the checker-boarding problem, meaning that the results have alternating zero and one elements like a checkerboard, which makes the structural element non-manufacturable \citep{diaz1995checkerboard}. Both occur  owing to numerical instabilities. We chose a sensitivity filter rather than a density filter to simplify the equations and produce the required results. To apply the sensitivity filter \citep{bendsoe2004topology}, sensitivities are modified using the following equation:
\begin{equation}
	\frac{\widehat{\partial c}}{\partial x_e}=\frac{1}{\max \left(\gamma, x_e\right) \sum_{j \in N_e} H_{e j}} \sum_{j \in N_e} H_{e j} \, x_j \, \frac{\partial c}{\partial x_j}.
\end{equation}

\noindent where $N_e$ is the set of elements $j$ that exists within the filter radius $r_{min}$, $\gamma=10^{-3}$ is a small positive number that is added to avoid division by zero, and $H_{e j}$ is a weight factor defined as follows:
\begin{equation}
	H_{e j}=\max \big( 0, r_{\min }-\Delta(e, j) \big).
\end{equation}

\noindent where $\Delta(e, j)$ is the distance between the center of element $e$ and the center of element $j$.

\subsection{Modelling Limitations}
This section discusses two design limitations\textcolor{red}{;} how initial discretization can affect results and how to cope with significantly high stress values in certain elements. Because stresses are localized, a mesh dependence test is necessary before relying on the results. Mesh dependency testing includes reducing the element size until the findings are no longer affected by the mesh size. If the elements are big, there will be a large number of intermediate elements with stress values less than the stress limit. Small components, on the other hand, will produce solid structural elements.

\noindent Another difficulty that might occur when modelling these types of problems is that certain parts in the vicinity of the load application might have a stress peak, which is not necessarily the case in actual applications since load is not always concentrated at a single location. This can be resolved by spreading the load among many nodes. Another option is to avoid adding the elements close to the load application point to the clusters while computing the $P$-norm value. This has no effect on the outcome, since the elements in the loading zone constitute a very small proportion of the total mesh elements.

\noindent Furthermore, it is typically preferable to normalize the stresses before assigning them to the MMA to minimize numerical inaccuracies during the optimization process, which might occur when all the $P$-norm values are assigned as constraints to the MMA.

\begin{figure}
	\centering
	\includegraphics[width=0.99\linewidth]{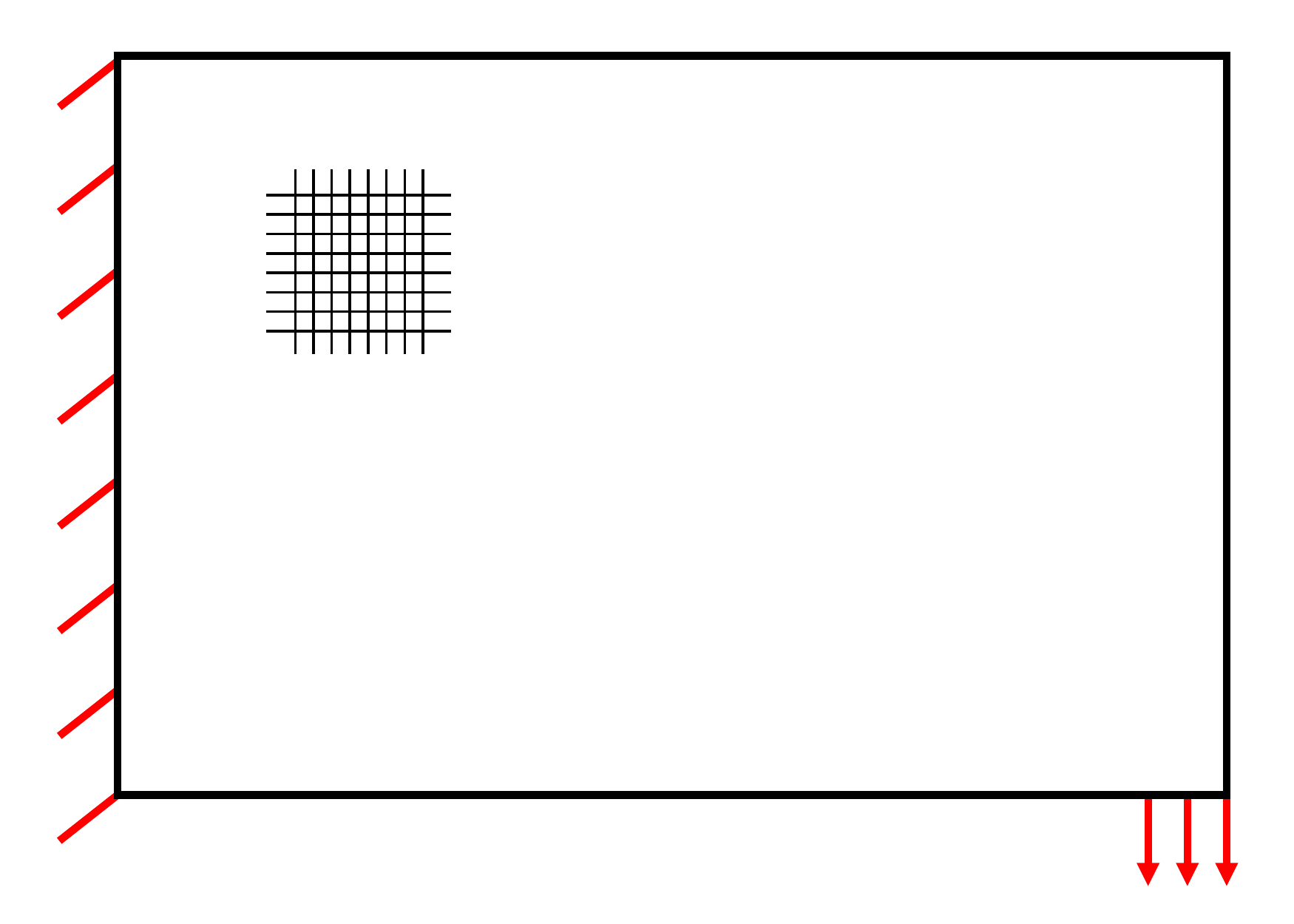}
	\caption{The cantilever beam is fully fixed from the left side. A distributed load on six adjacent nodes is applied at the bottom-right corner.}
	\label{fig:cant_beam_diag}
\end{figure}

\section{Numerical Cases}
\subsection{Case Study 1: Validation of the Method}

The first case study demonstrates the effectiveness of the stress constraints and how they can affect the final design. A compliance minimization problem subjected to a volume constraint is solved twice, first without adding the stress constraints. The second time, the stress constraints are set to a value lower than the maximum stresses calculated in the first case to ensure the constraints are active during the solution process. Mathematically, the first case can be written as Eq. (\ref{eq:general_eqtn}) while setting $V_f = 0.25$ and without the stress constraints. The second case can be represented by the same Eq. (\ref{eq:general_eqtn}) while setting $V_f = 0.25$, $\sigma^T_1 = \sigma^C_1 = 60$ kPa, and $\sigma^T_2 = \sigma^C_2 =$ 20 kPa. Both problems start from the same initial conditions of $\mathbf{\rho}=1$ and $\mathbf{\theta}= - 0.1$ rad.

The geometry is shown in Fig. \ref{fig:cant_beam_diag} with the load value set to 60 kN distributed along six consecutive nodes, with 10 kN per node.
The design domain was discretized by 60 elements in $x$ and 40 elements in $y$; a total of 2400 elements. The stress cluster used had 240 stress evaluation points. One cluster was used for $\sigma_1^{PN}$ and one for $\sigma_2^{PN}$. For all the case studies, the material properties used are for epoxy glass and defined according to \cite{kaw2005mechanics} as: $E_1 = 38.6$ GPa, $E_2 = 8.27$ GPa, $G_{12} =$ 4.14 GPa, $\nu_{12} =$ 0.27, and $\nu_{21} =$ 0.0578. The penalization power value is set to 3 and $P$-norm coefficient is set to 8 except for case study 4 where the effect of different $P$-norm coefficients is studied. The convergence criteria is set such that the algorithm is stopped when the maximum change in density is lower than 0.001.

\begin{figure}
	\centering
	\subfloat[Without Stress Constraints.]{\includegraphics[width=0.99\columnwidth]{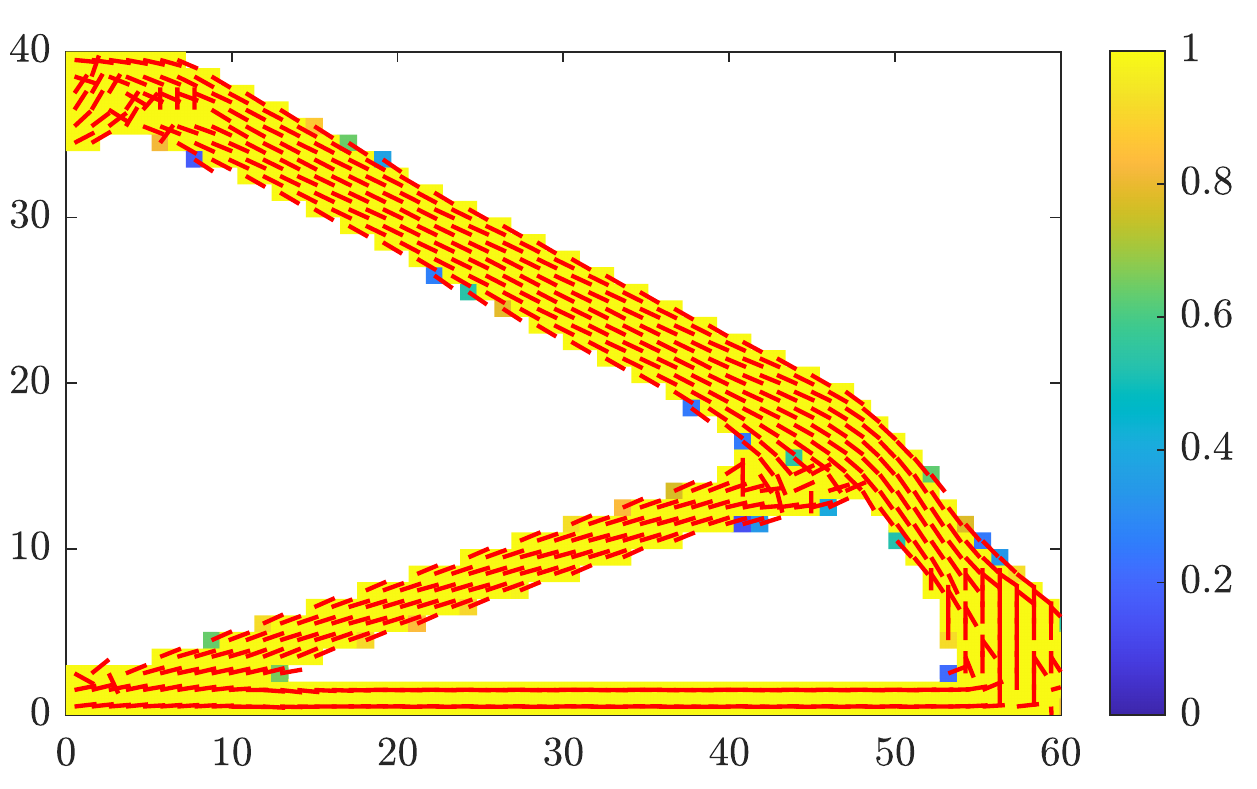}} \\
	\subfloat[With Stress Constraints.]{\includegraphics[width=0.99\columnwidth]{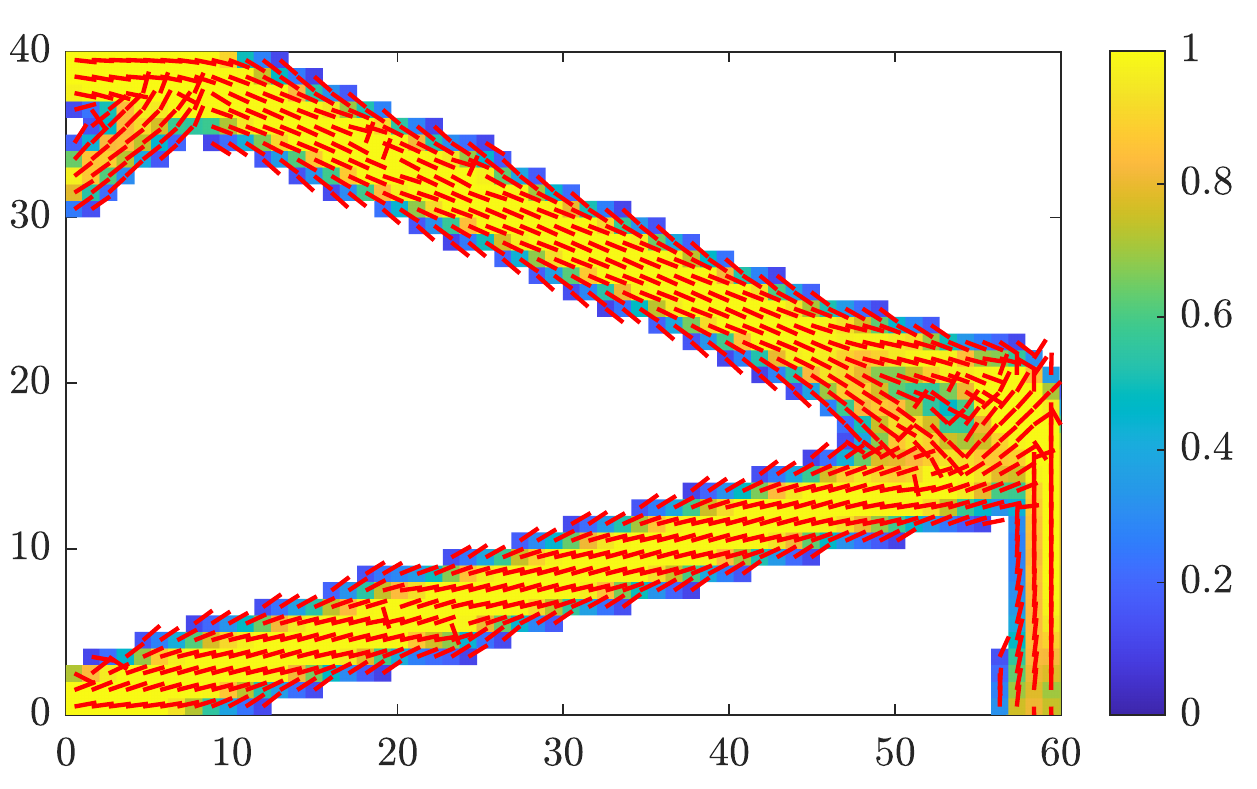}}
	\caption{Optimized topology and fiber orientation for case study 1. The colorbar represents the density while the red dashes represent the fiber orientation.}
	\label{fig:ch5p1top}
\end{figure}

\begin{figure}
\centering
	\subfloat[Without stress constraints.]{\includegraphics[width=0.99\columnwidth]{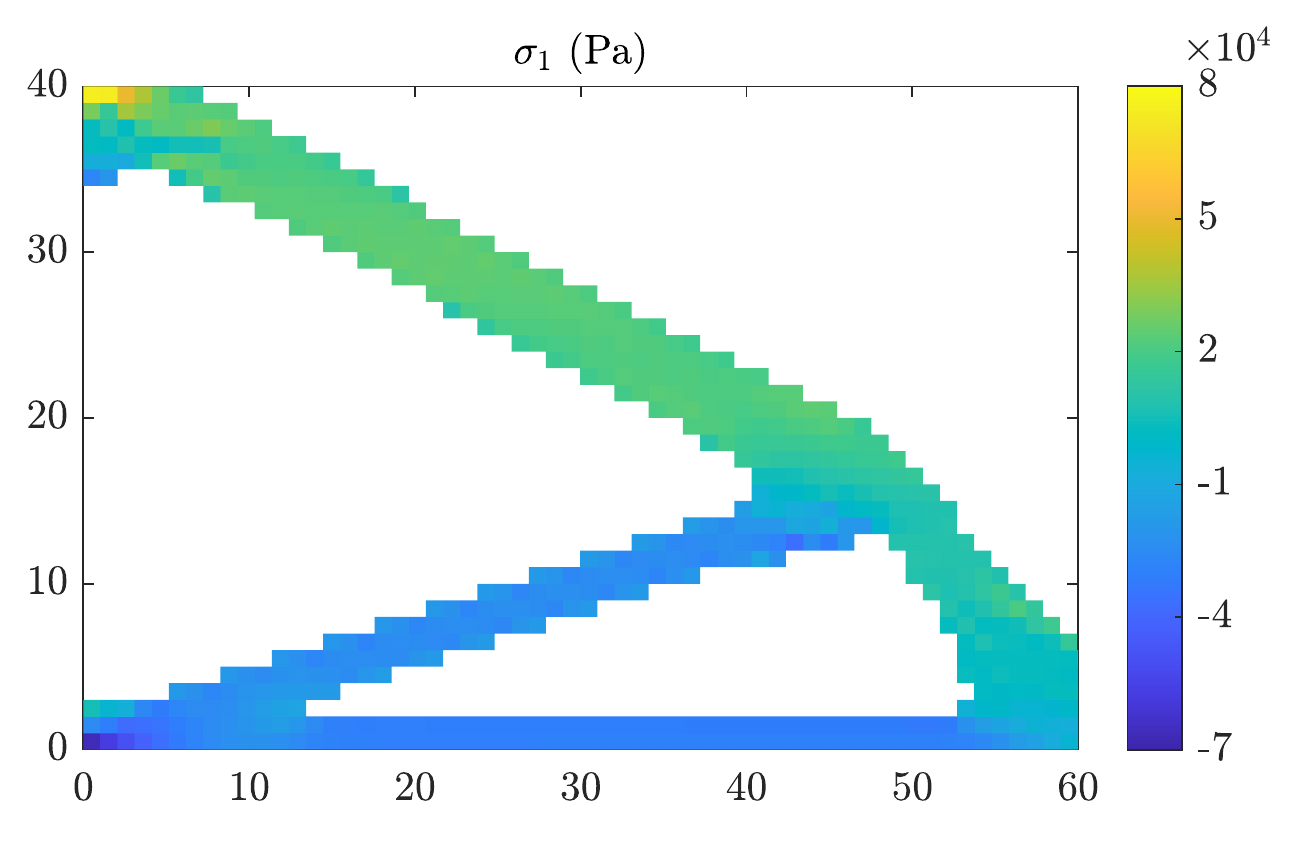}} \\
	\subfloat[With stress constraints.]{\includegraphics[width=0.99\columnwidth]{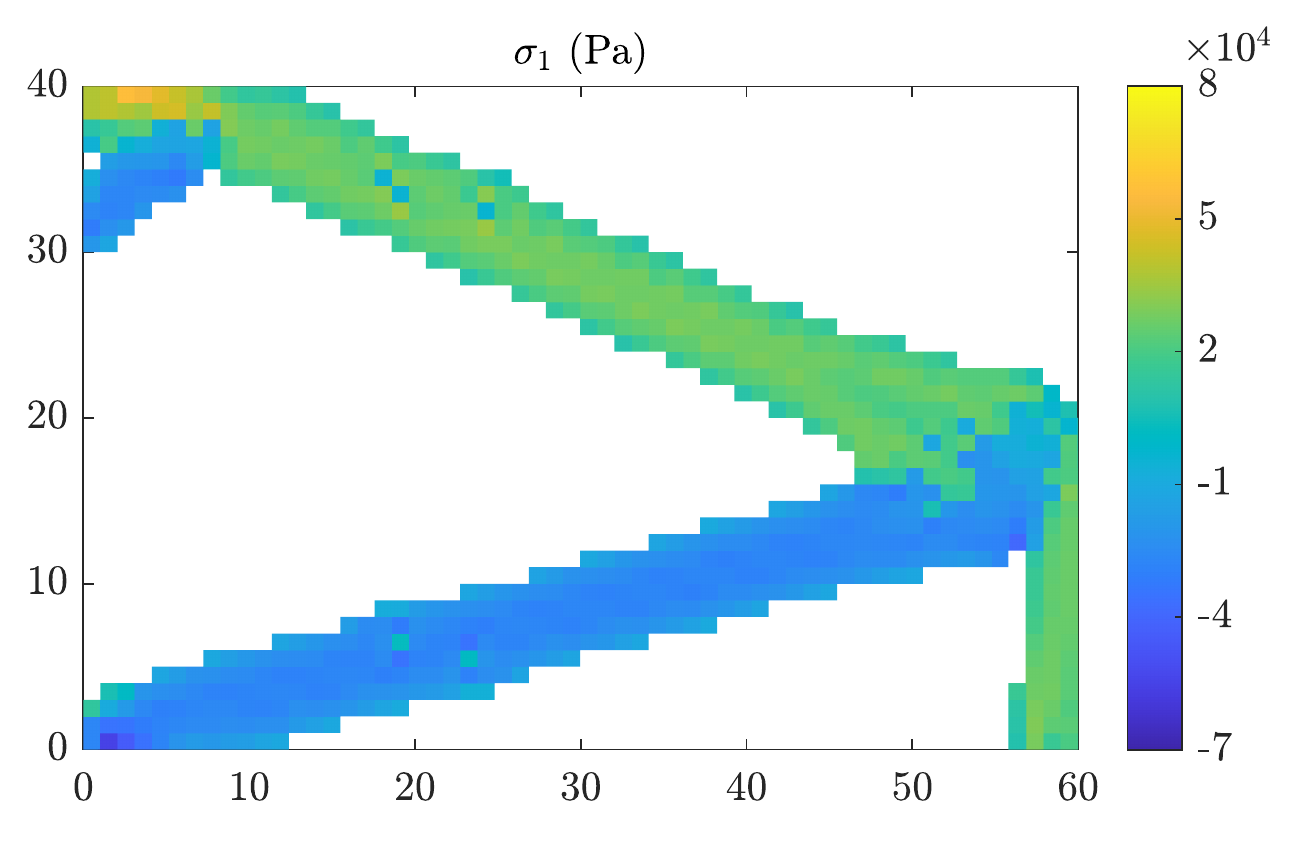}}
	\caption{Stresses along the fibers for case study 1.}
	\label{fig:ch5p1sx}
\end{figure}

\begin{figure}
	\centering
	\subfloat[Without stress constraints.]{\includegraphics[width=0.99\columnwidth]{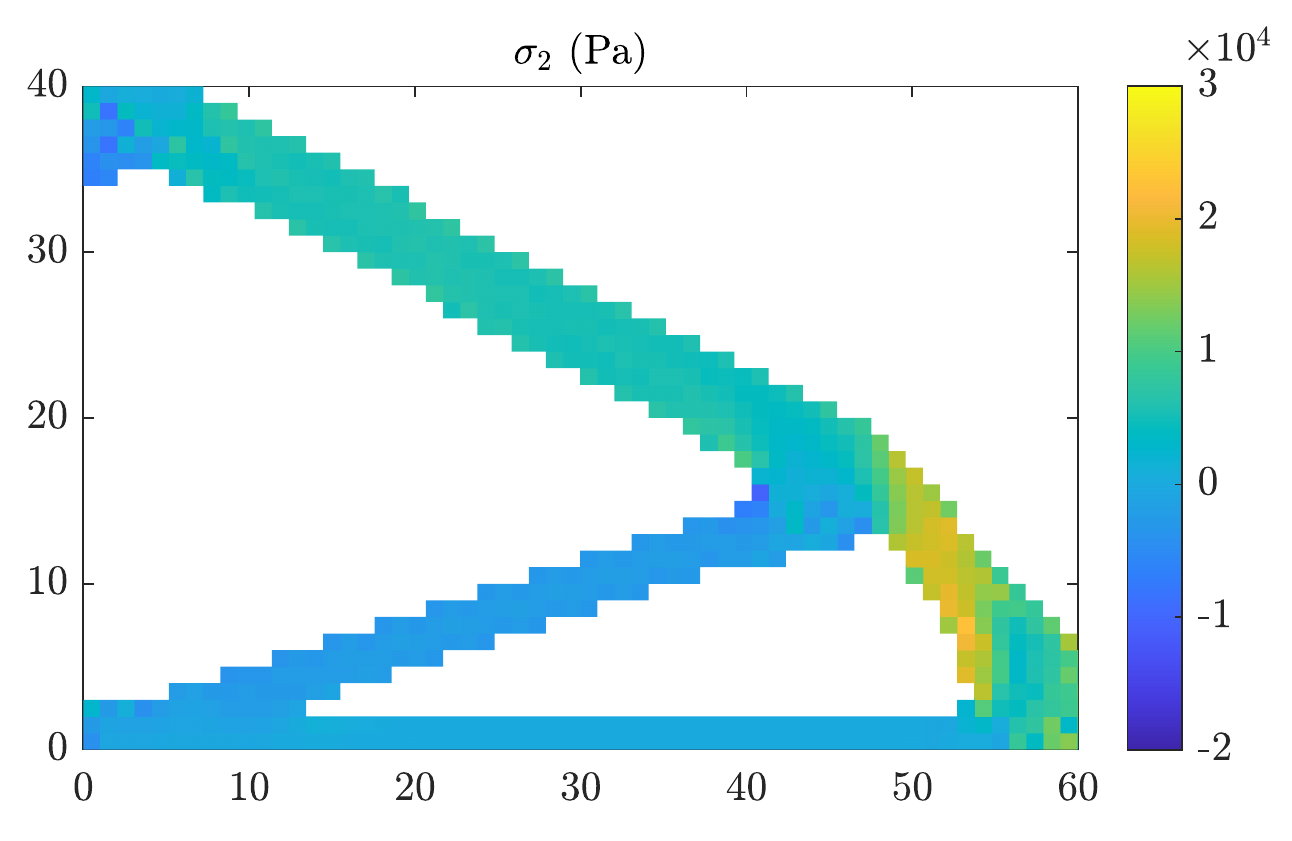}} \\
	\subfloat[With stress constraints.]{\includegraphics[width=0.99\columnwidth]{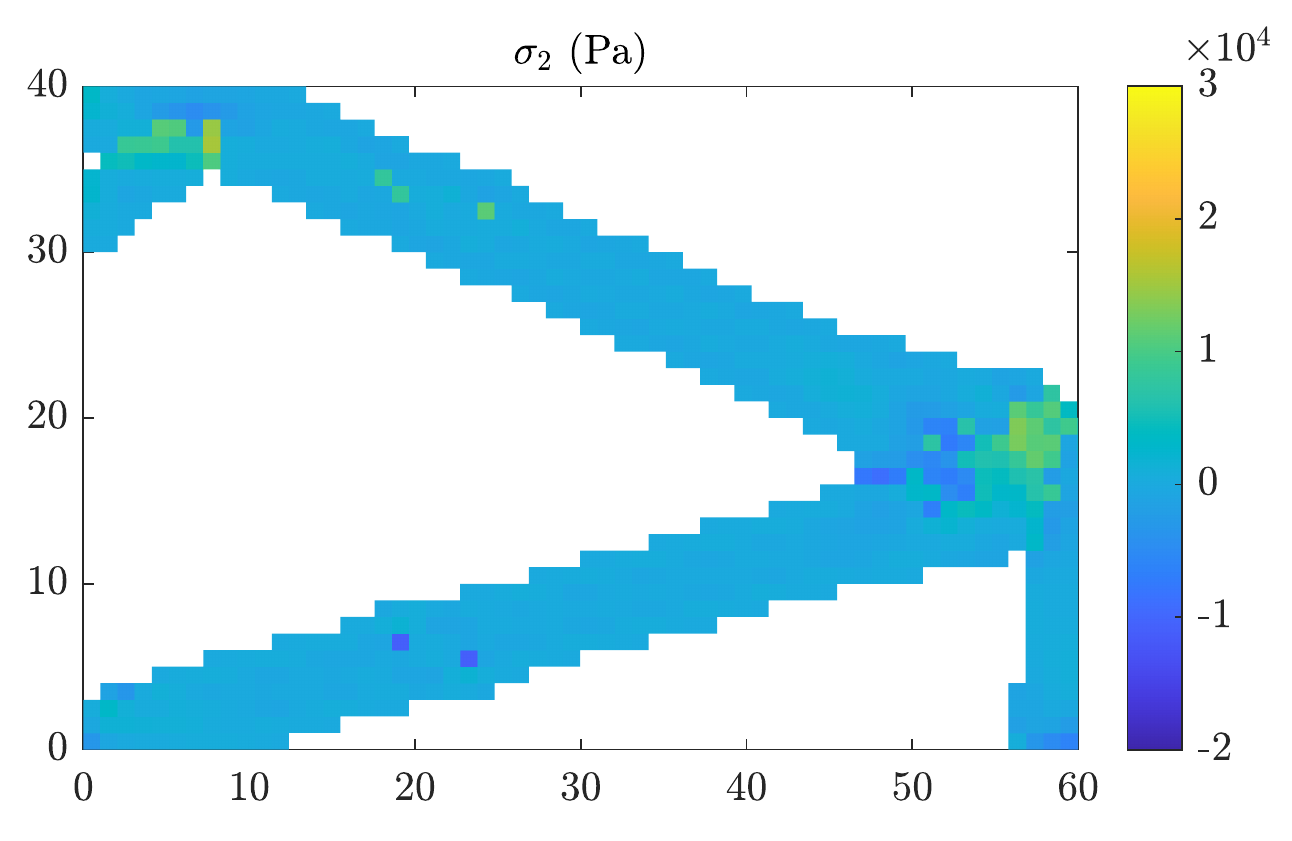}}
	\caption{Stresses perpendicular to the fibers for case study 1.}
	\label{fig:ch5p1sy}
\end{figure}

The results show a compliance of 9.0542 N$\cdot$m for the problem without stress constraints and a maximum stress value along the fibers of 74.88 kPa in tension and 66.45 kPa in compression. For stresses perpendicular to the fibers, the stress values reached 22.4 kPa in tension and 10.2 kPa in compression. However, after setting the stress constraint, a new topology was designed by the software, which yielded a compliance value of 9.77 N$\cdot$m and stresses that fell within the constraints. In observing the stress-constrained topology, it is clear that a novel topology was obtained, where most of the stresses were carried by the bottom-right member. Hence, the stresses on the top-left corner decreased to reach the constraint values. The new topology reduced the stresses along the fibers to 56.4 and 55.1 kPa in tension and compression, respectively. For the stress values perpendicular to the fibers, they were reduced to 15.4 and 11 kPa in tension and compression, respectively. Figures \ref{fig:ch5p1sx} and \ref{fig:ch5p1sy} illustrate the stress distribution of both topologies. For this problem, both problems showed that fibers should be aligned with the structure contours to obtain minimum compliance as shown in Fig. \ref{fig:ch5p1top}.

\subsection{Case Study 2: Effect of Number of Stress Evaluation Points on Final Topology}

\begin{figure}
	\centering
	\includegraphics[width=0.99\columnwidth]{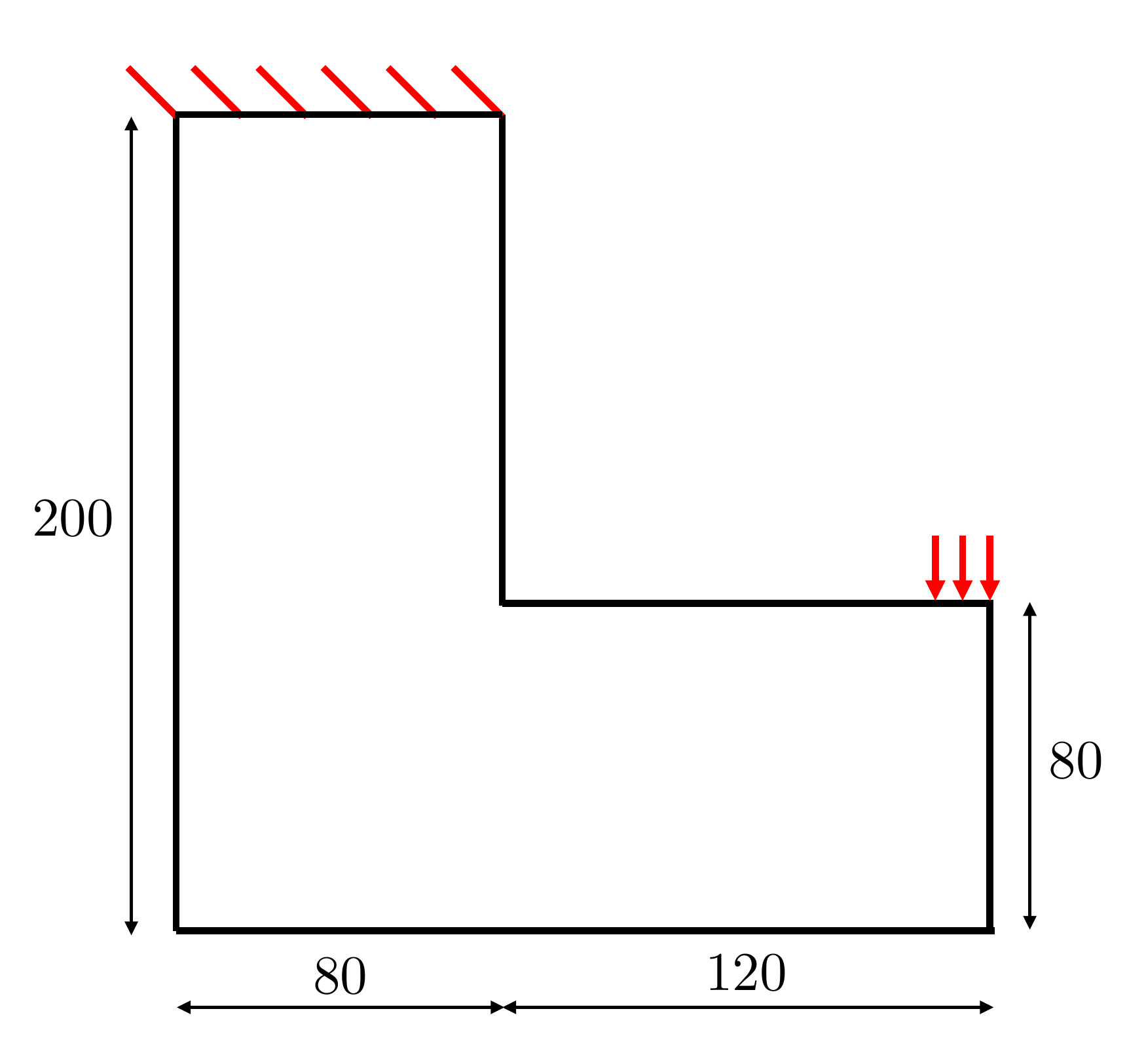}
	\caption{L-bracket structure subjected to 60 kN loading and fixed from the upper-left corner.}
	\label{fig:L}
\end{figure}

Case study 2 evaluates the effect of the number of stress evaluation points included in the cluster used as a constraint for each of the material local axes on the accuracy of the result and convergence scheme. Two studies are run; one with 80 stress evaluation points (i.e. finite elements) per cluster and one with 40 stress evaluation points per cluster. The L-bracket design domain shown in Fig. \ref{fig:L} is used with a load of 60 kN and the indicated fixed boundary condition and the material is epoxy glass. Both problems can be represented by Eq. \ref{eq:general_eqtn} while setting $V_f = 0.25$, $\sigma^T_1 = \sigma^C_1 = 5 $ kPa, and $\sigma^T_2 = \sigma^C_2 = 4 $ kPa. Both problems start from the same initial conditions of $\mathbf{\rho}=1$ and $\mathbf{\theta}= -0.1 $ rad.

\begin{figure}
	\centering
	\subfloat[40 stress evaluation points in the stress cluster.]{\includegraphics[width=0.99\columnwidth]{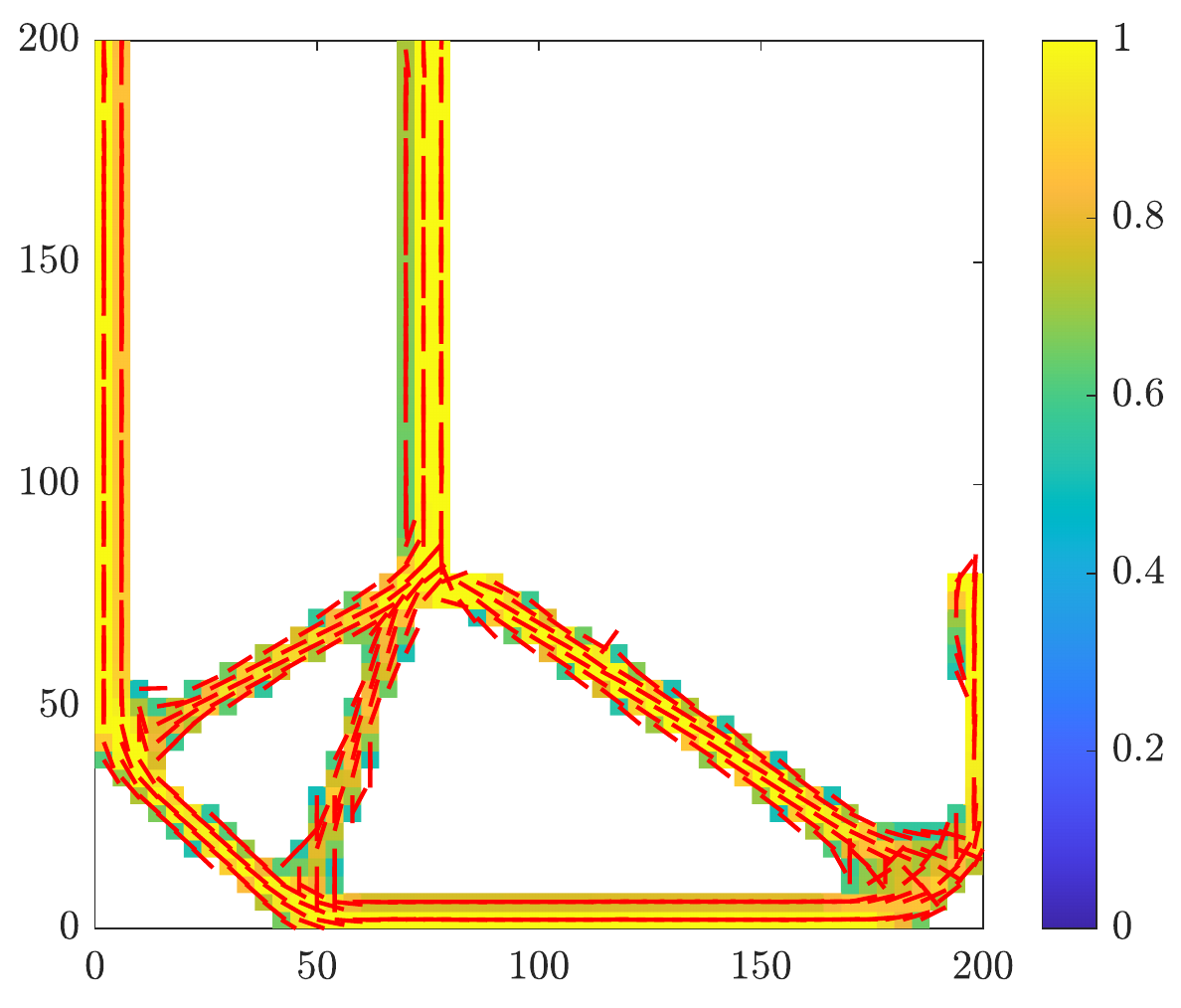}} \\
	\subfloat[80 stress evaluation points in the stress cluster.]{\includegraphics[width=0.99\columnwidth]{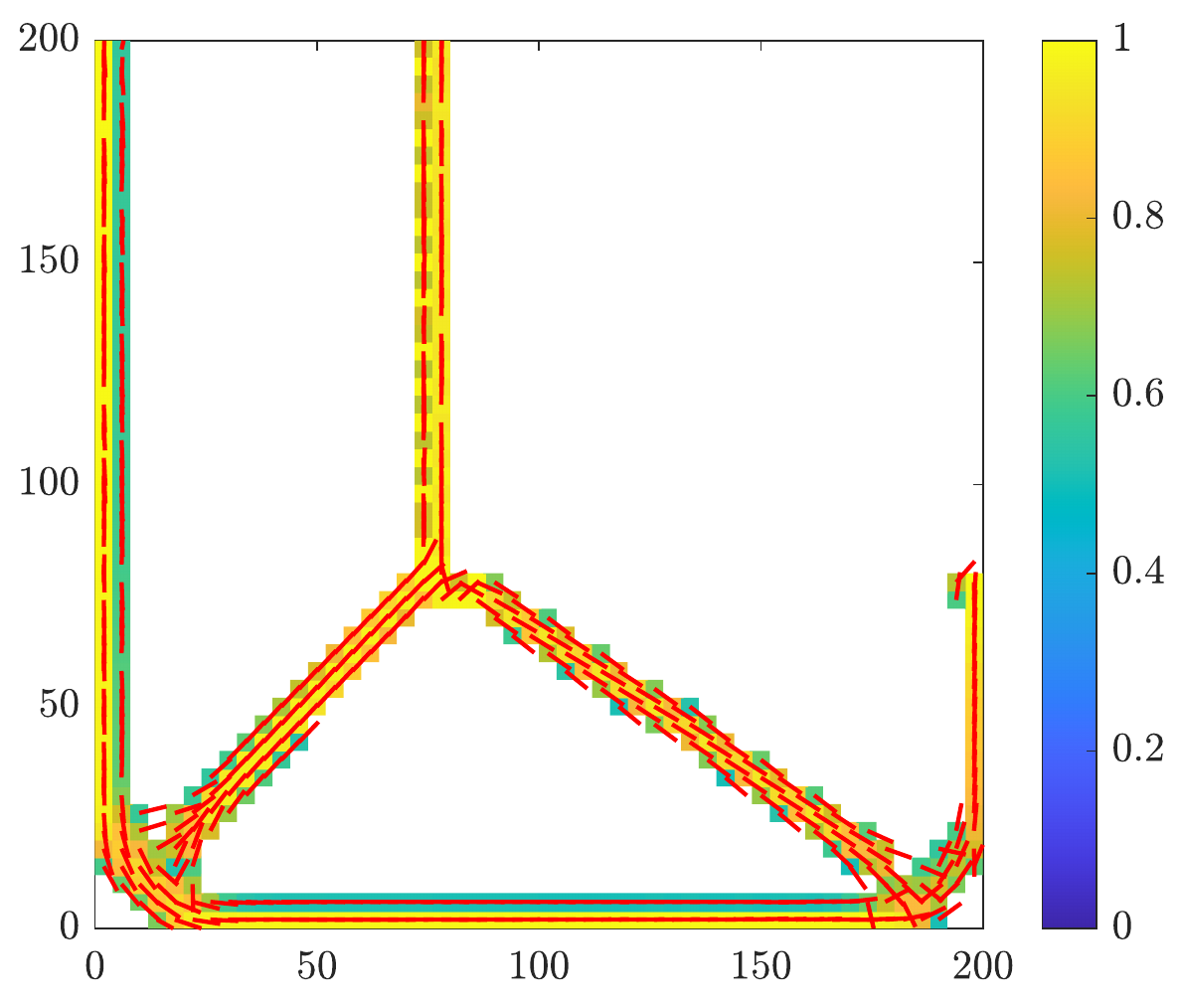}}
	\caption{Final topology and fiber orientation for case study 2.}
	\label{fig:ch5p2top}
\end{figure}

\begin{figure}
\centering
	\subfloat[40 stress evaluation points in the stress cluster. \label{fig:40along}]{\includegraphics[width=0.99\columnwidth]{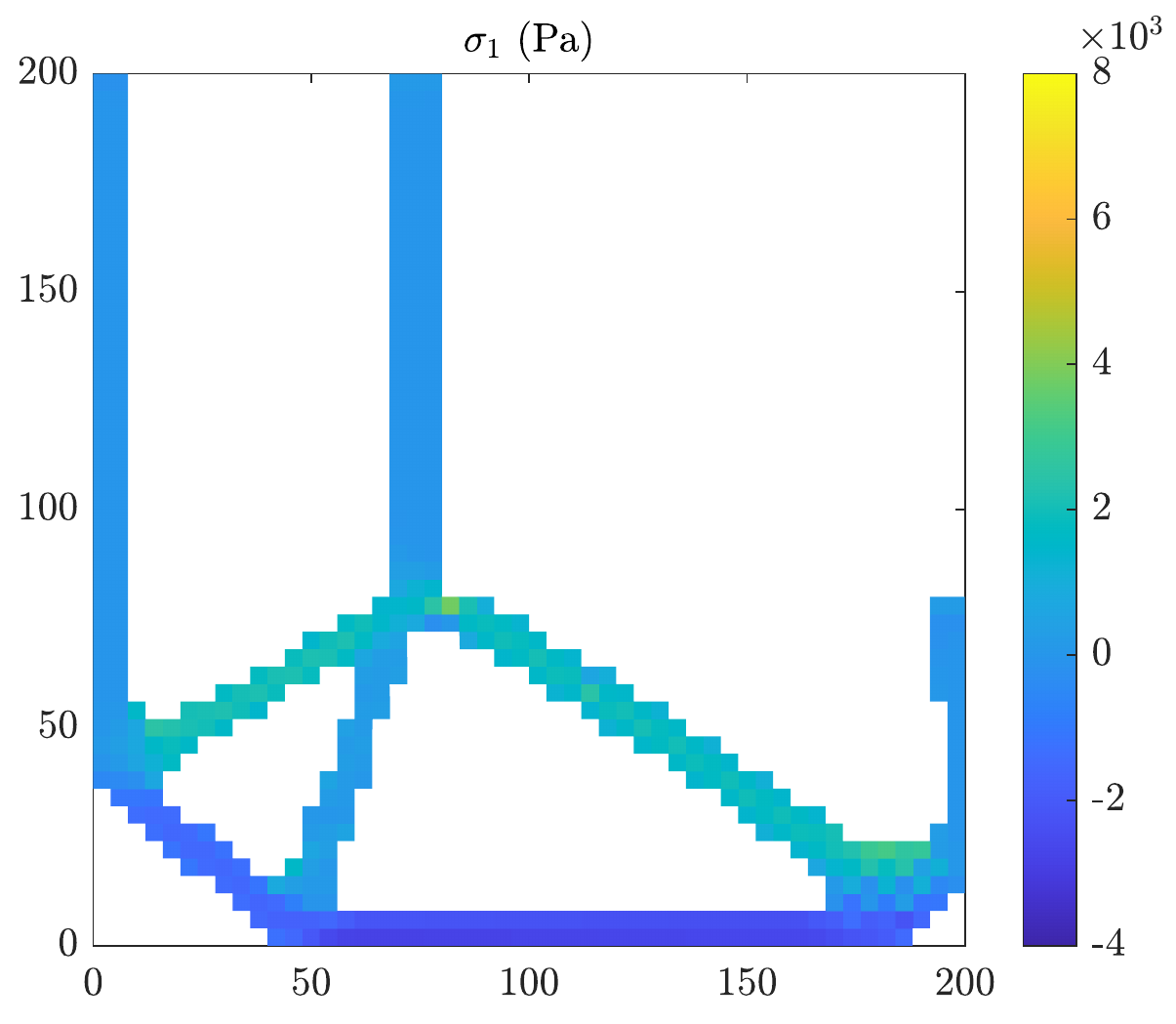}} \\
	\subfloat[80 stress evaluation points in the stress cluster. \label{fig:80along}]{\includegraphics[width=0.99\columnwidth]{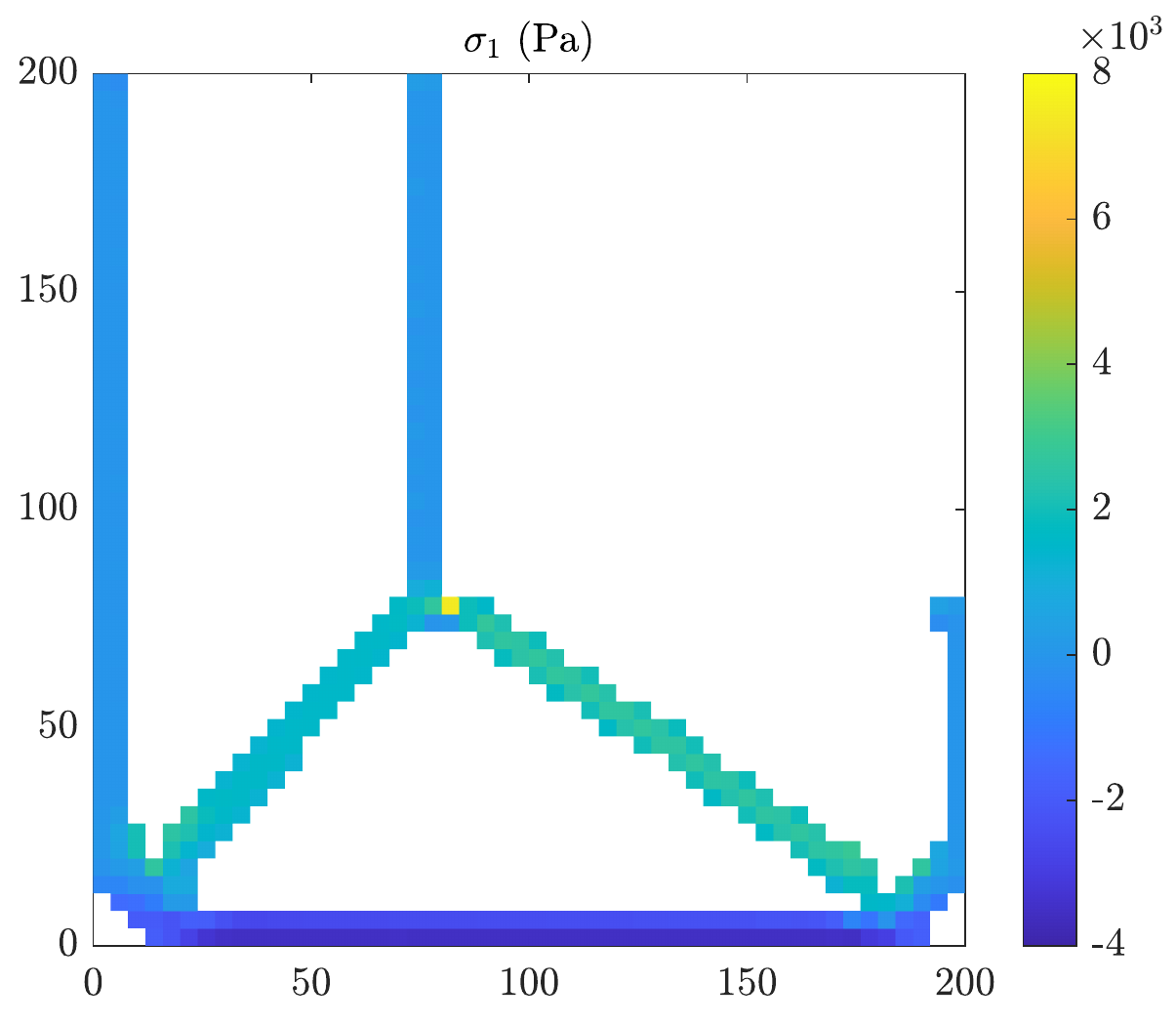}}
	\caption{Stresses along the fibers for case study 2.}
	\label{fig:ch5p2sx}
\end{figure}

\begin{figure}
\centering
	\subfloat[40 stress evaluation points in the stress cluster. \label{fig:40perp}]{\includegraphics[width=0.99\columnwidth]{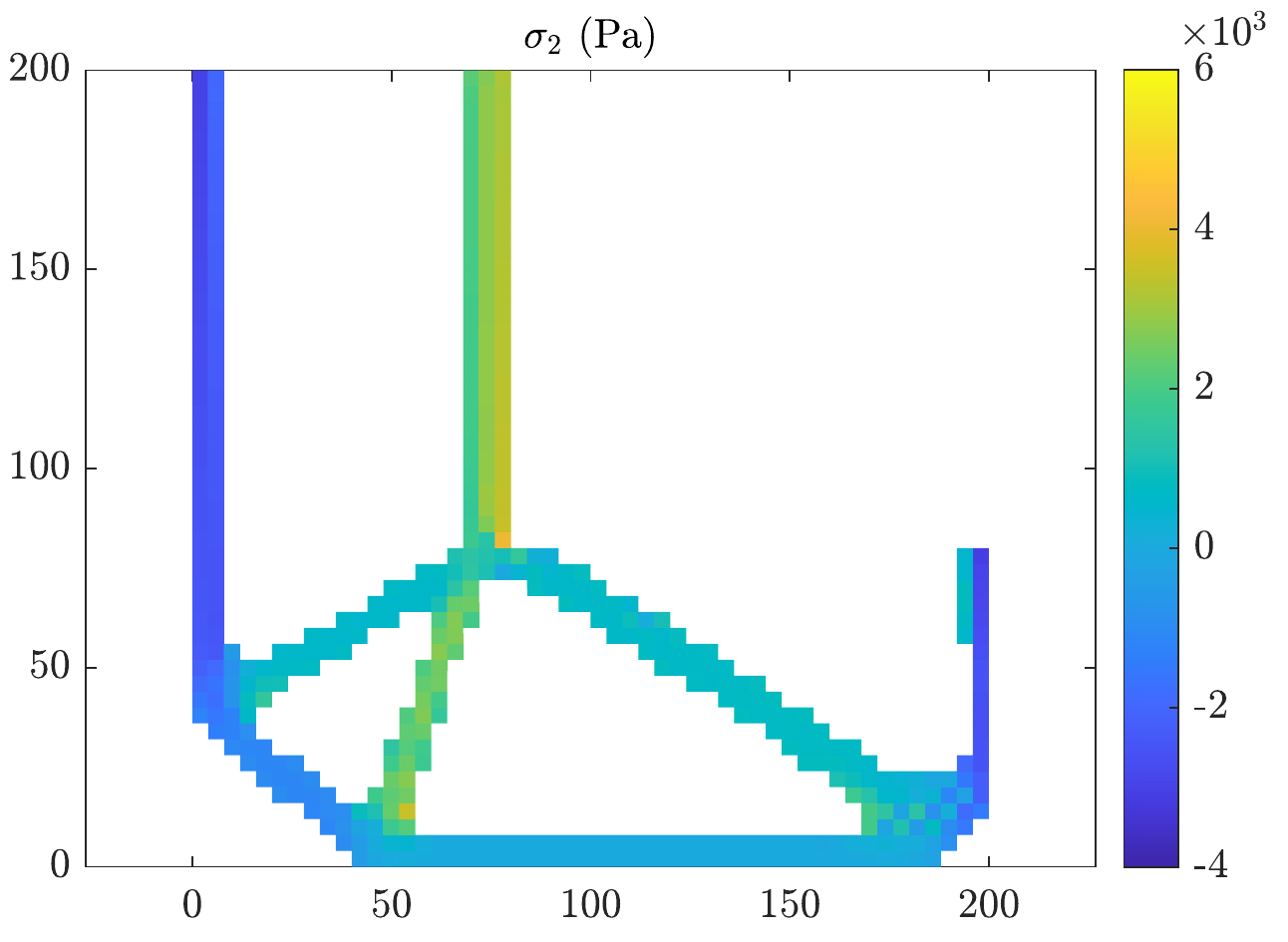}} \\
	\subfloat[80 stress evaluation points in the stress cluster. \label{fig:80perp}]{\includegraphics[width=0.99\columnwidth]{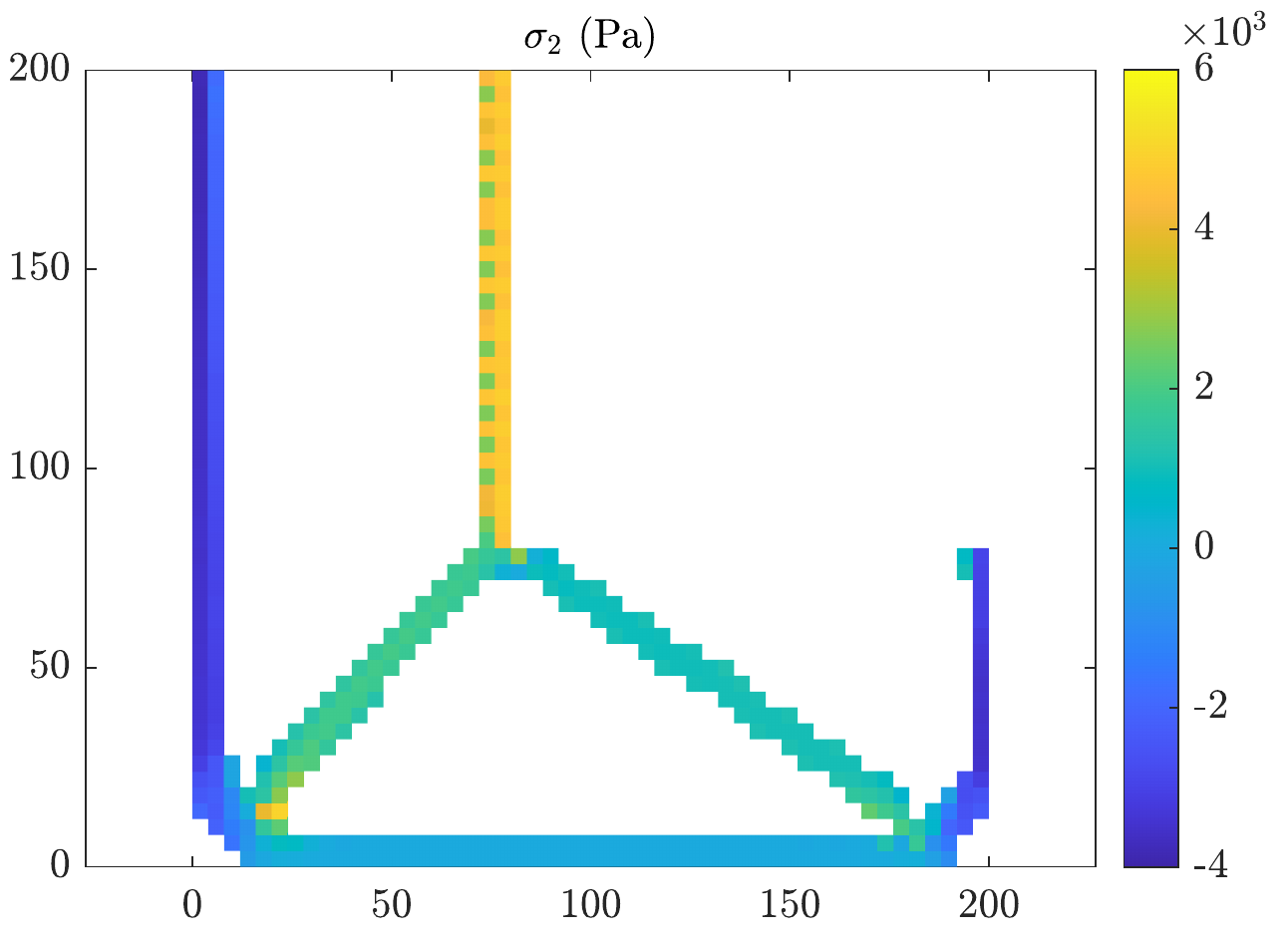}}
	\caption{Stresses perpendicular to the fibers for case study 2.}
	\label{fig:ch5p2sy}
\end{figure}

\begin{figure}
    \centering
	\subfloat[40 stress evaluation points in the stress cluster.]{\includegraphics[width=0.99\columnwidth]{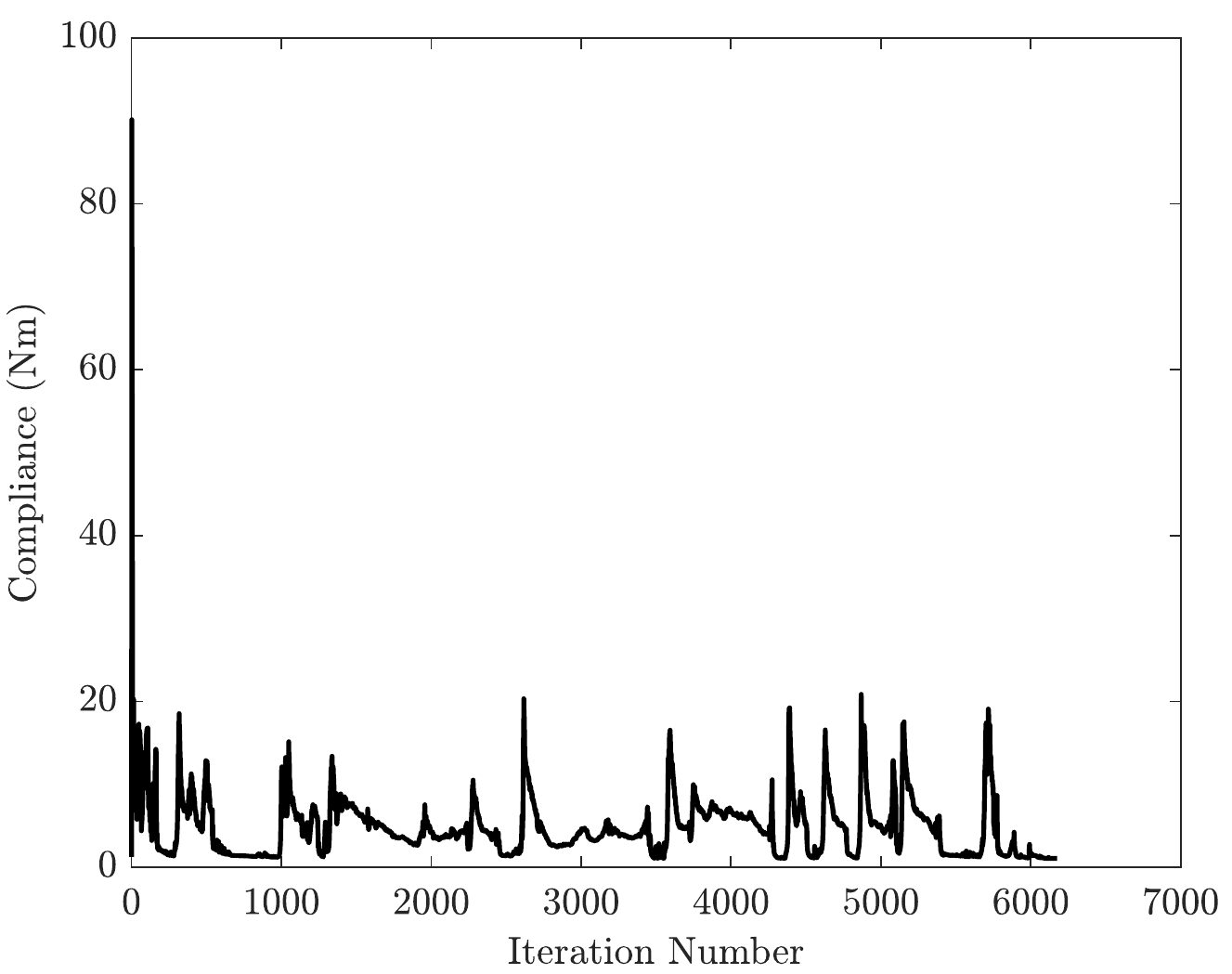}} \\
	\subfloat[80 stress evaluation points in the stress cluster.]{\includegraphics[width=0.99\columnwidth]{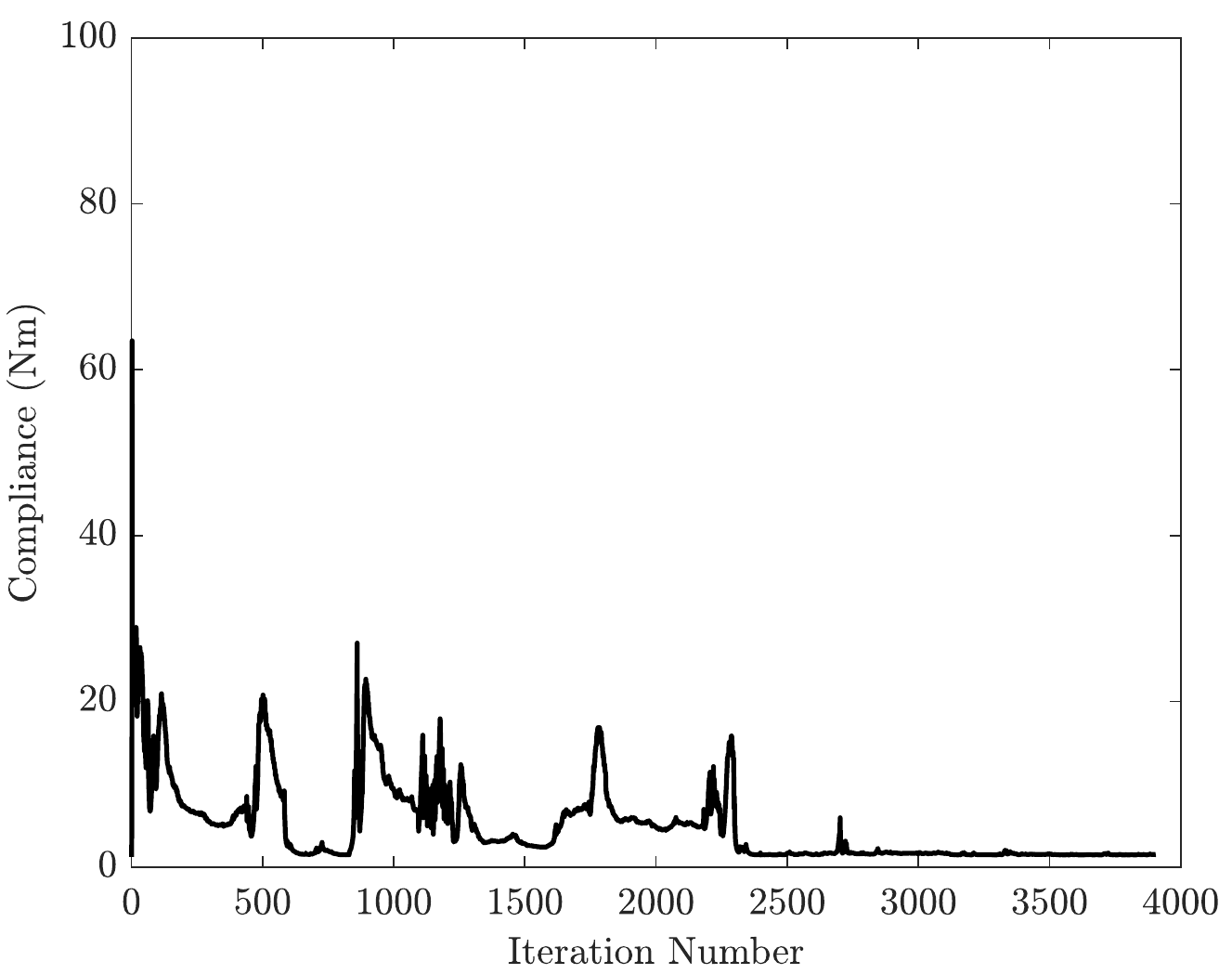}}
	\caption{Convergence scheme of the objective functions for case study 2.}
	\label{fig:ch5p2conv}
\end{figure}

The results show new topologies compared to previous work done in the field as shown in Fig. \ref{fig:ch5p2top}. Using 80 stress evaluation points resulted in a more flat L-bracket than the one generated with 40 evaluation points. Moreover, adding too many stress evaluation points to the cluster may decrease the effectiveness of the algorithm, as the $P$-norm value - and hence the constraint value - will not be a good indicator of the stress values throughout the domain, as indicated in Table \ref{tbl:clusters}. This scenario was clear in the case with 80 points, as stresses did not fall within the constraint range and some outliers existed. On the other hand, when using 40 evaluation points, all the stresses fell within the constraint range as shown in Figs. \ref{fig:ch5p2sx} and \ref{fig:ch5p2sy}. Moreover, compliance values were found to be 1.0583 and 1.5923 N$\cdot$m for the 40 evaluation points and the 80 evaluation points, respectively. This finding confirms that the lower number of evaluation points is more efficient. The case with a higher number of evaluation points was better in one respect, which is the convergence speed, as it converged in only 3905 iterations, whereas in the former case, 6176 iterations were required, as shown in Fig. \ref{fig:ch5p2conv}. However, accurate results, not faster convergence, is the main aim of this study. High non-linearity is observed in Fig. \ref{fig:ch5p2conv}. Oscillations are more pronounced in the case of 40 evaluation points per cluster and that is the anticipated result as locality of constraints impairs convergence (c.f. \citet[p.~9]{Abdelhamid2021}).

\begingroup
\renewcommand{\arraystretch}{1.25} 
\begin{table*}[hbt!]
	\centering
	\begin{tabular}{ m{8cm} | m{1.5cm}      m{1.5cm}  m{1.5cm}}
		\hline
		No. of clusters in each direction & 1 & 1 & 2  \\ \hline
		No. of stress evaluation points per cluster & 40 & 80 & 40
		\\ \hline 
		Max. stress along the fibers (kPa)
		&
		4.5
		&
		7.4
		& 
		5.2 
		\\ \hline
		Max. stress perpendicular to the fibers (kPa) & 4.01 & 5.1 & 4.2
		\\ \hline
		Compliance (N$\cdot$m) & 1.05 &  1.592 & 1.49
		\\ \hline
		$\sigma_1^{PN1}$ (kPa) & 5.04 &  6.23 & 4.62
		\\ \hline
		$\sigma_2^{PN1}$ (kPa) & 4.2 & 4.5 & 3.84
		\\ \hline
		$\sigma_1^{PN2}$ (kPa) & N/A &  N/A & 3.86
		\\ \hline
		$\sigma_2^{PN2}$ (kPa) & N/A &  N/A & 3.58
		\\ \hline
		No. of iterations & 6176 & 3905 & 2515
		\\ \hline
	\end{tabular}
	\caption{Comparison of different stress clustering techniques.}
	\label{tbl:clusters}
\end{table*}
\endgroup

\subsection{Case Study 3: Adding One More Stress Cluster in Each Direction}

Case study 2 demonstrated that decreasing the number of stress evaluation points in the maximum stress cluster leads to more accurate results but also a longer convergence time, on the order of ~6000 iterations. One solution to this problem is to constrain the first two clusters in each evaluation direction, instead of taking only one cluster. By doing so, the solver is constraining 80 stress evaluation points at the same time but dividing them into two clusters. That proved to be the most efficient way of dealing with a problem like the L-bracket, which is famous for producing different topologies depending on the constraints. In this case, we will redo the L-bracket example from case study 2, but instead constrain two clusters in each direction.\\

\begin{figure}
    \centering
	\subfloat[Topology and fiber orientation. \label{fig:2cltrstop}]{\includegraphics[width=0.99\columnwidth]{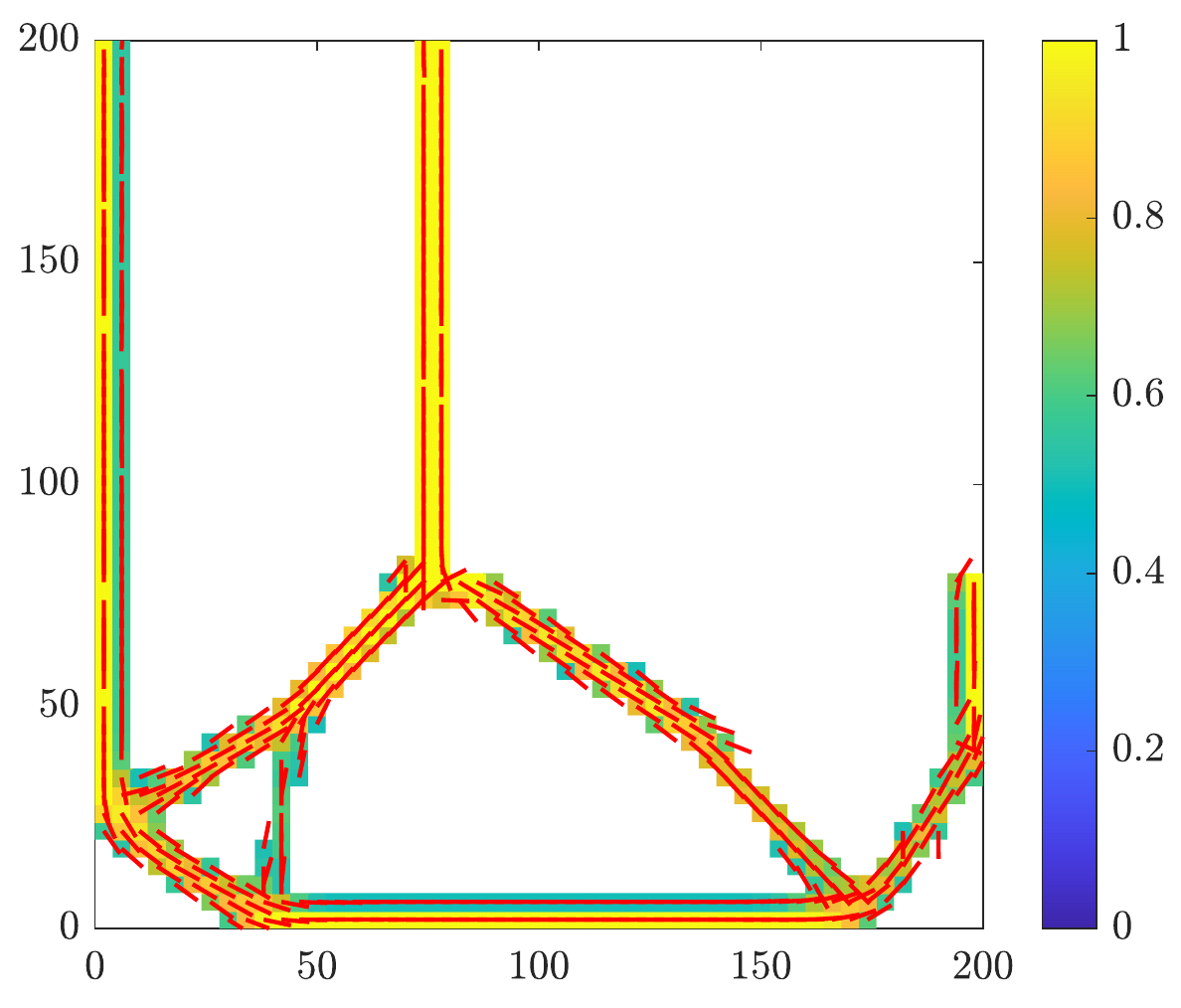} } \\
	\subfloat[Convergence scheme.]{\includegraphics[width=0.99\columnwidth]{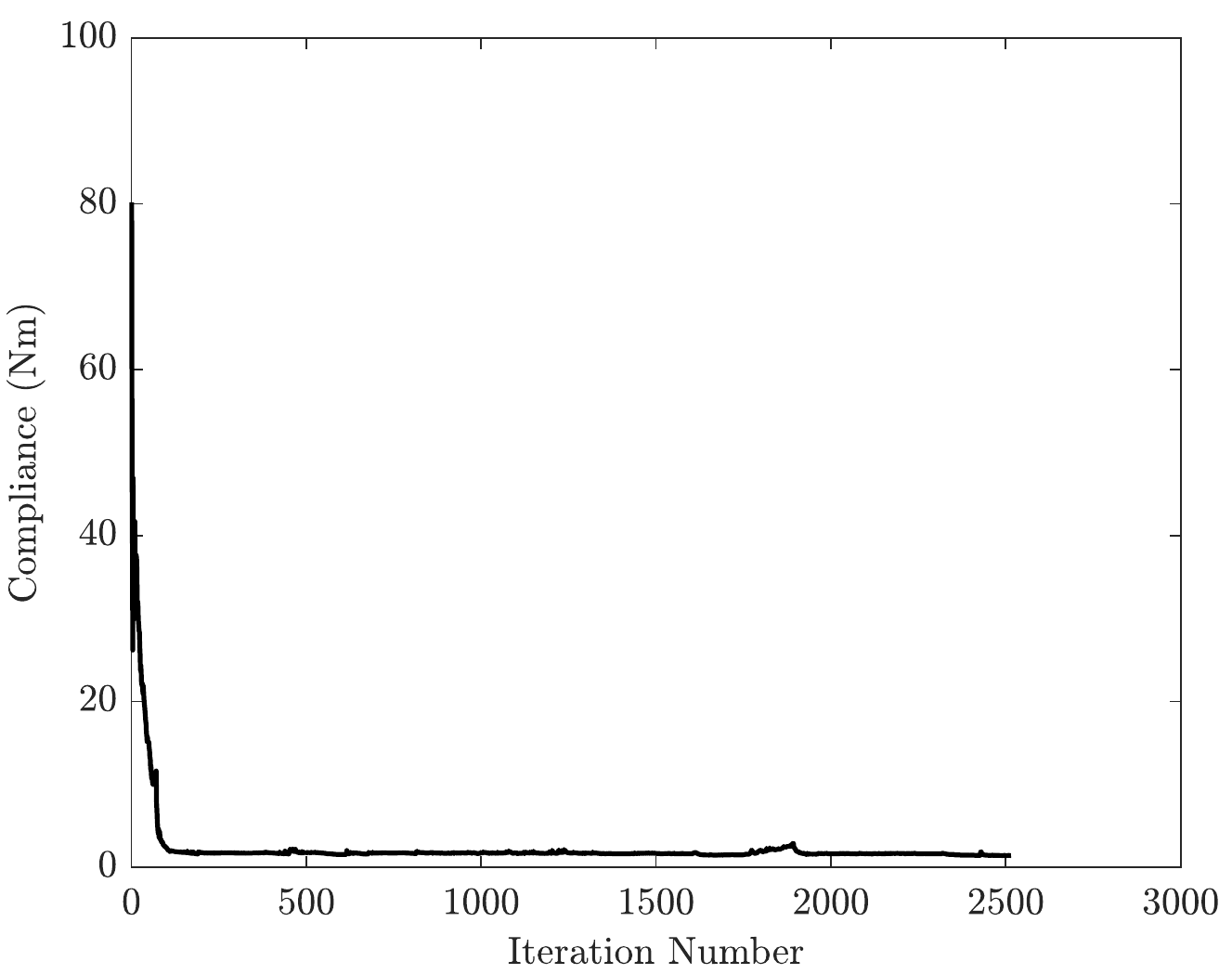}}
	\caption{Results of case study 3 using two stress clusters in each of the material coordinate directions.}
\end{figure}

\noindent The results show that this method is the most efficient, as it produced a topology with more solid elements under the loading point, as seen in Fig. \ref{fig:2cltrstop}. Moreover, all stress values fell within the constraint limit and converged in 2515 iterations, which is a faster convergence compared with the previous two cases. The compliance value was 1.49 N$\cdot$m, and the constraints were satisfied as shown by the $P$-norm values in Table \ref{tbl:clusters}.


\begin{figure*}
    \centering
	\subfloat[Stresses along the fibers.]{\includegraphics[width=0.99\columnwidth]{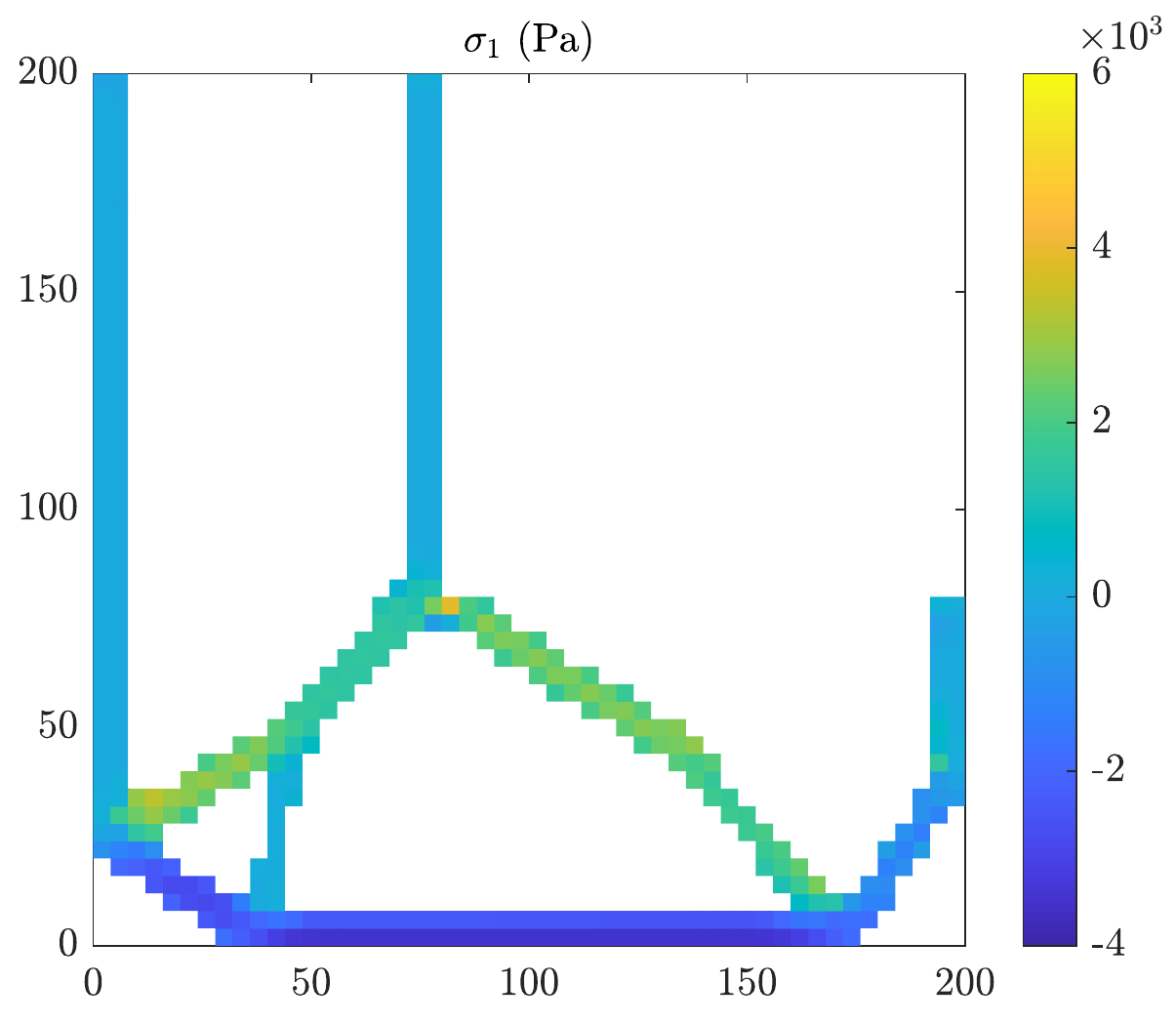}}
	\subfloat[Stresses perpendicular to the fibers.]{\includegraphics[width=0.99\columnwidth]{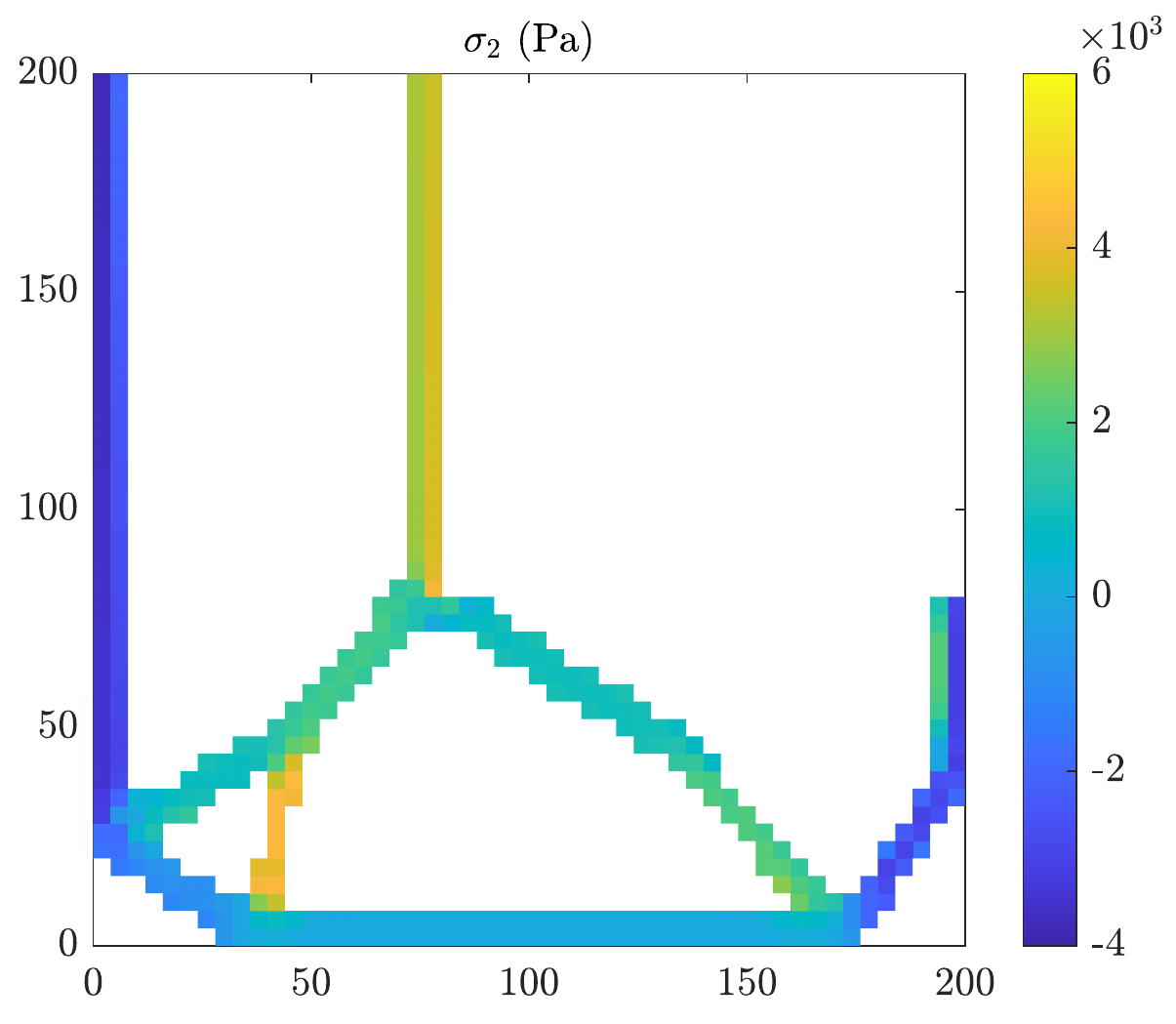}}
	\caption{Stress values for case study 3 when using two stress clusters in each of the material coordinate directions.}
	\label{fig:ch5p4top}
\end{figure*}

\subsection{Case Study 4: Comparing $P$-values} \label{pnorm}
The $P$-norm method is so called because the arithmetic series of Eq. \ref{eq:clustering} is raised to the power of $P$ \citep{holmberg2013stress}. This number is chosen based on trial and error. From literature, it is usually set to $P = 8$ \citep{holmberg2013stress}. However, that value from literature corresponds to an isotropic material and constrains all the stress clusters. Case study 4 tests the effect of the $P$-value on stress values generated in the principal material coordinates of an orthotropic material, namely epoxy glass. This study is performed on the same geometry, constraints, and boundary and loading conditions as those used in case study 1. However, the initial conditions are set as $\rho=$ 0.25 and $\theta=$ 0.1 rad.

\begin{figure*}
	\centering
	\subfloat[$P=4$]{\includegraphics[width=0.99\columnwidth]{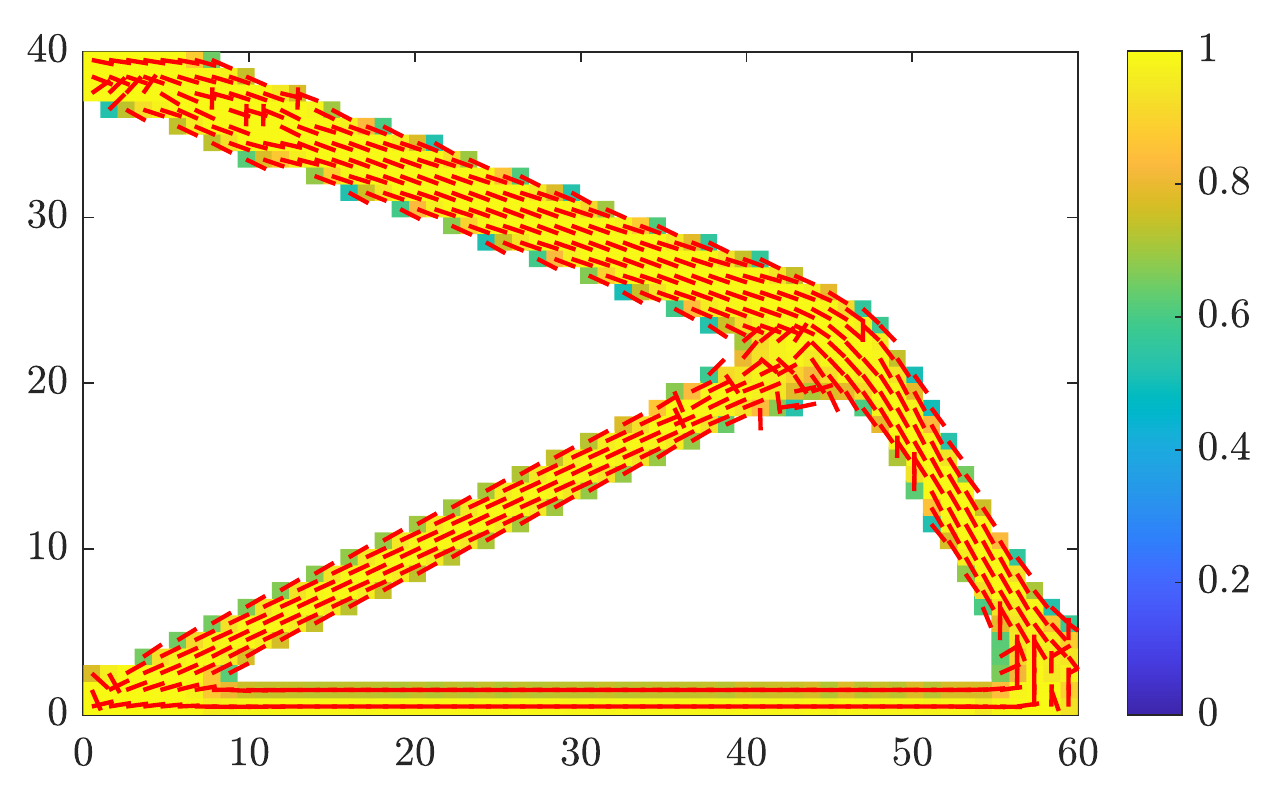}}
	\subfloat[$P=6$]{\includegraphics[width=0.99\columnwidth]{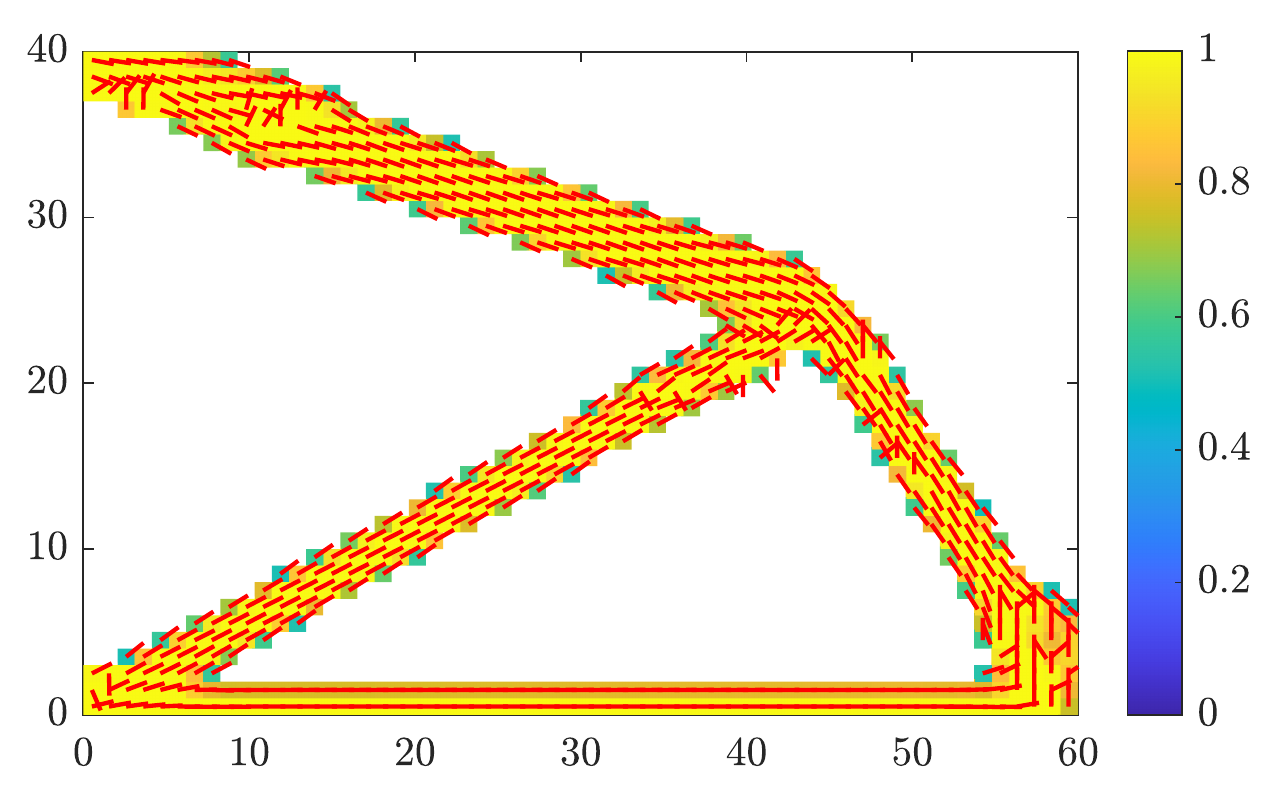}} \\
	\subfloat[$P=8$]{\includegraphics[width=0.99\columnwidth]{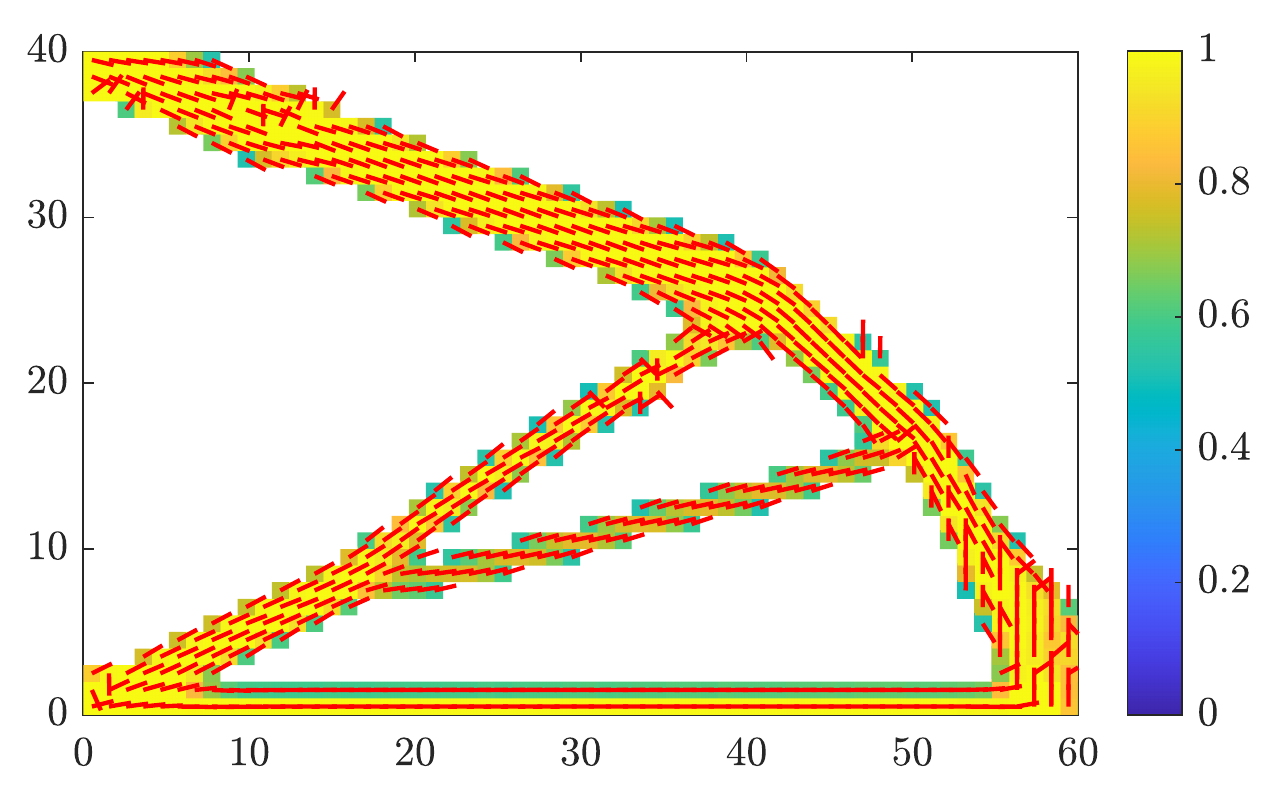}}
	\subfloat[$P=10$]{\includegraphics[width=0.99\columnwidth]{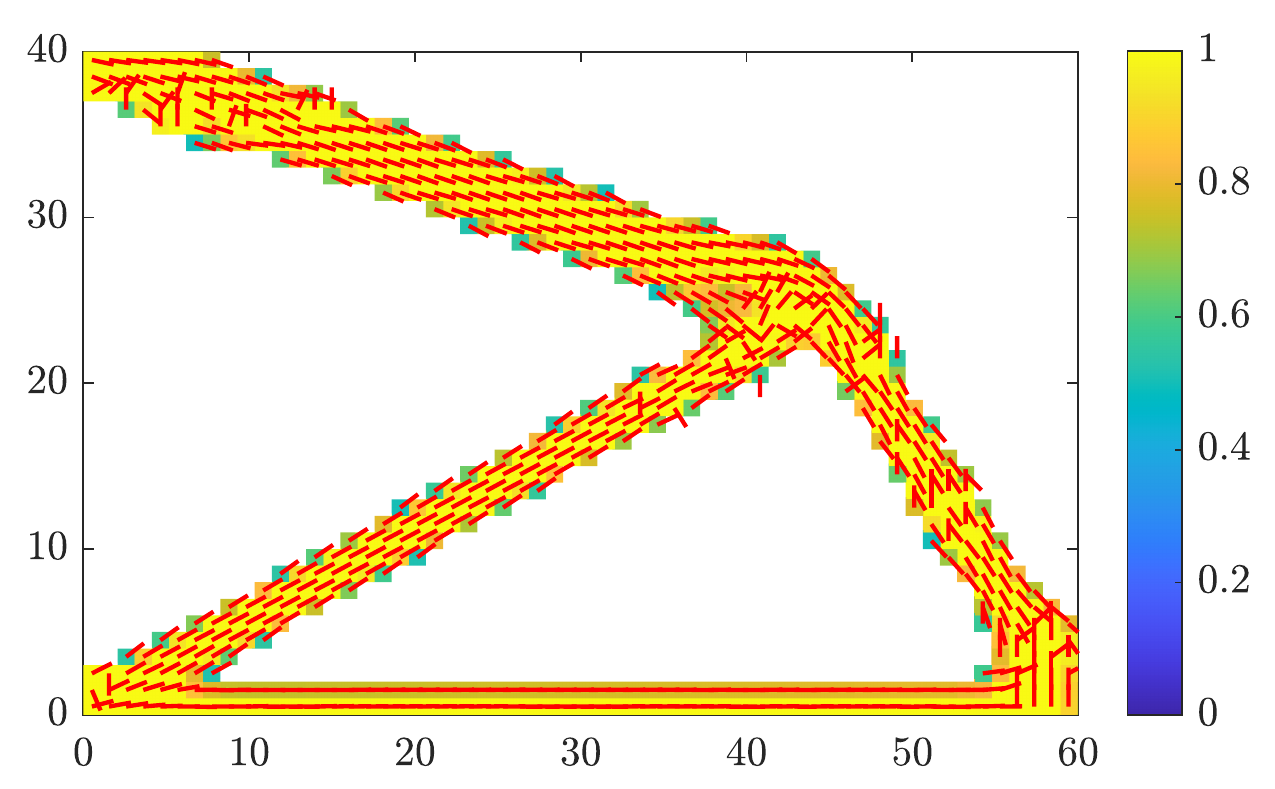}}
	\caption{Effect of $P$-value on topology and fiber orientation (case study 4).}
	\label{fig:ch5p3top}
\end{figure*}

In observing the results shown in Table \ref{tbl:P}, it seems that all $P$ values keep the stress values within the specified constraints. A $P$-value of 8 seems to be the best compromise considering it gives the lowest stress values at a relatively moderate total number of iterations. It's also worth noting that the 8 $P$ value gives a significantly different topology from the other $P$ values as seen in Fig. \ref{fig:ch5p3top}. These results are in good agreement with the discussion on the value of $P$ for isotropic materials in \citep[p.~611]{Le2010}\footnote{The only difference from the results in \cite{Le2010} is the number of iterations of $P=6$ which should have been somewhere between those of $P=4$ and $P=8$. From our numerical experiments with slightly different stress constraint values, the iteration numbers seem to align with the trend observed in \cite{Le2010}.}. As a final note, the maximum stress values along the fibers reported in Table \ref{tbl:P} exclude two elements with overshooting stress values located in the left top and left bottom corners (i.e. the ends of the fixed constraint).

\begingroup
\renewcommand{\arraystretch}{1.25} 
\begin{table*}
	\centering
	\begin{tabular}{ m{8cm} | m{1cm}  m{1cm}  m{1cm} m{1cm} }
		\hline
		$P$-value & 4 & 6 & 8 & 10 \\ \hline
		Max. stress along the fibers (kPa)
		& 58.00 & 57.00 & 53.51 & 58.14
		\\ \hline
		Max. stress perpendicular to the fibers (kPa) & 12.50 & 12.40 & 10.82 & 13.63
		\\ \hline
		Compliance (N$\cdot$m) & 8.45 & 8.40 & 8.94
		& 8.51 \\ \hline
		No. of iterations & 140 & 224 & 188
		& 274 \\ \hline
		No. of stress evaluation points per cluster & 120 & 120 & 120
		& 120 \\ \hline
	\end{tabular}
	\caption{Effect of $P$-value on the results of case study 4 with $\sigma^T_1$ = 60 kPa and $\sigma^T_2$ = 25 kPa.}
	\label{tbl:P}
\end{table*}
\endgroup

\begin{figure*}
	\centering
	\subfloat[$P=4$]{\includegraphics[width=0.99\columnwidth]{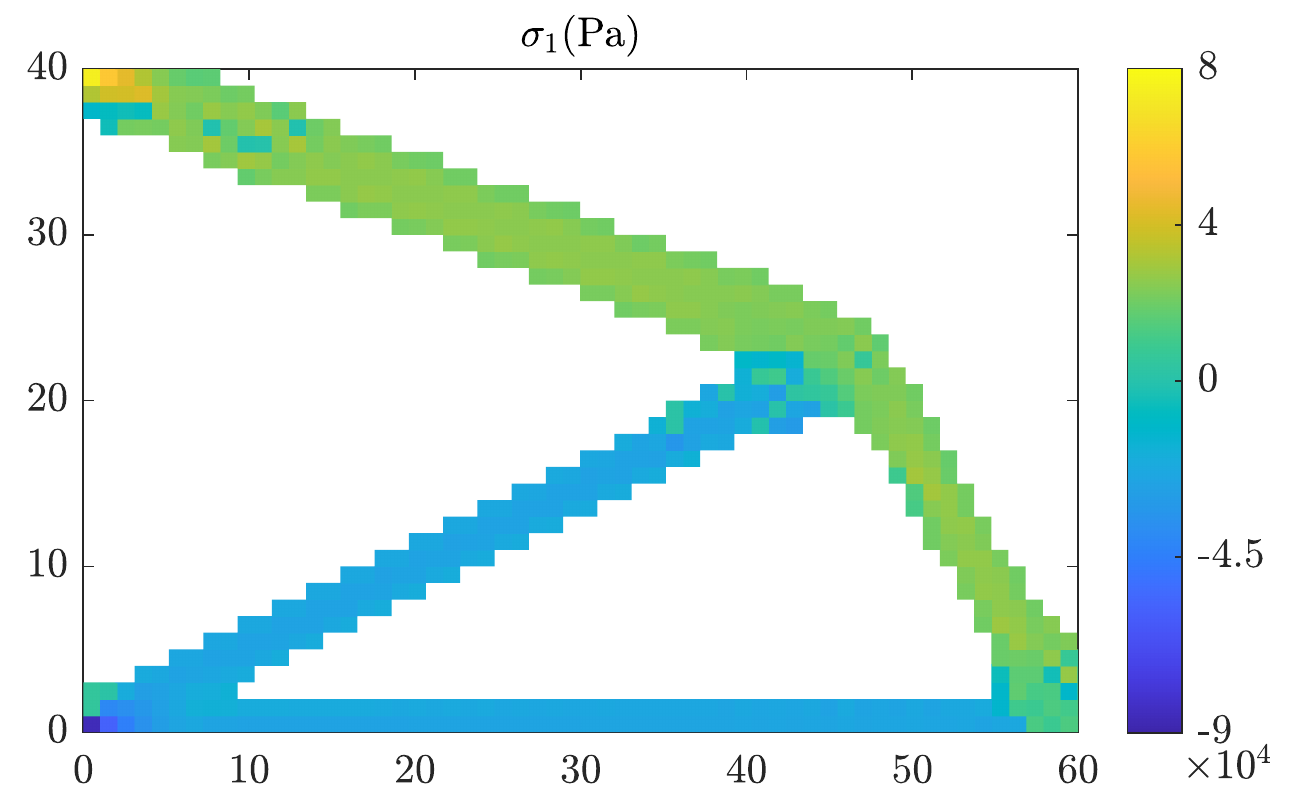}}
	\label{fig:4sx} 
	\subfloat[$P=6$]{\includegraphics[width=0.99\columnwidth]{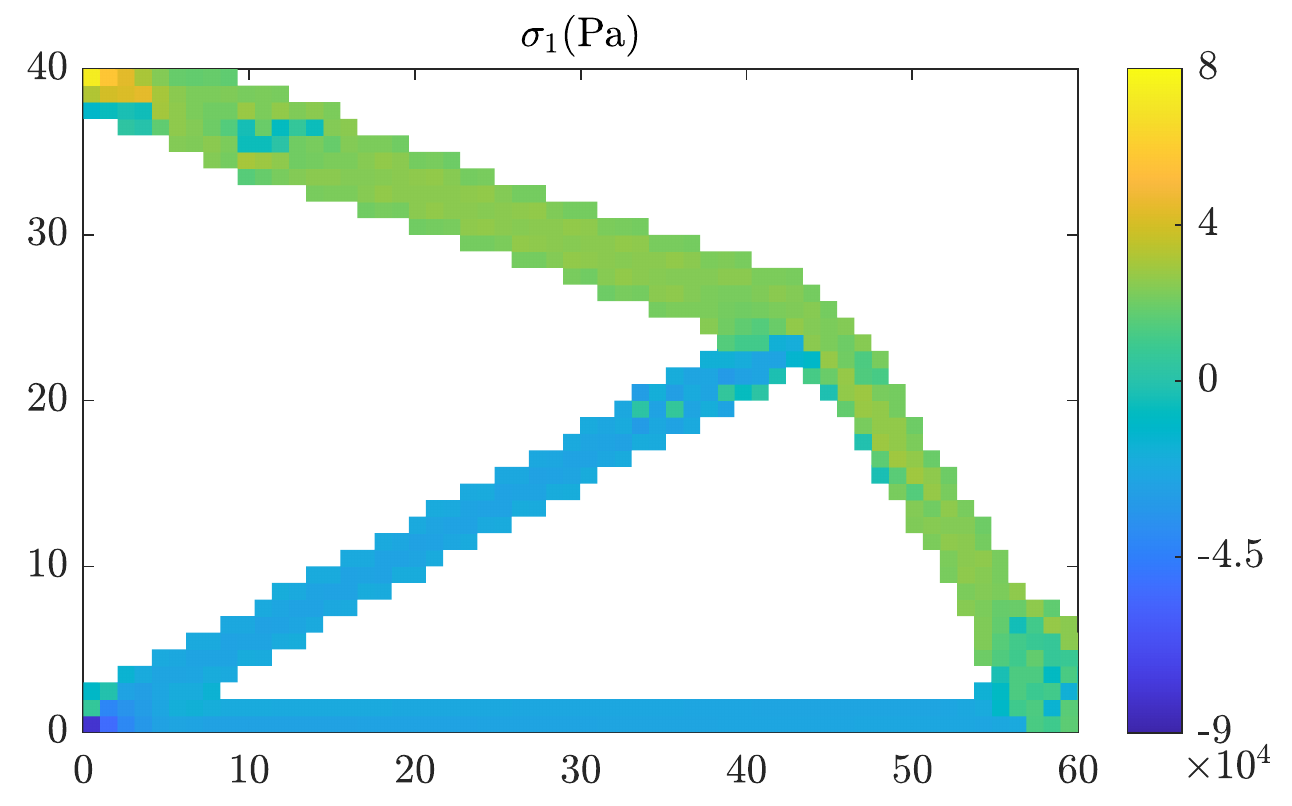}}
	\label{fig:6sx} \\
	\subfloat[$P=8$]{\includegraphics[width=0.99\columnwidth]{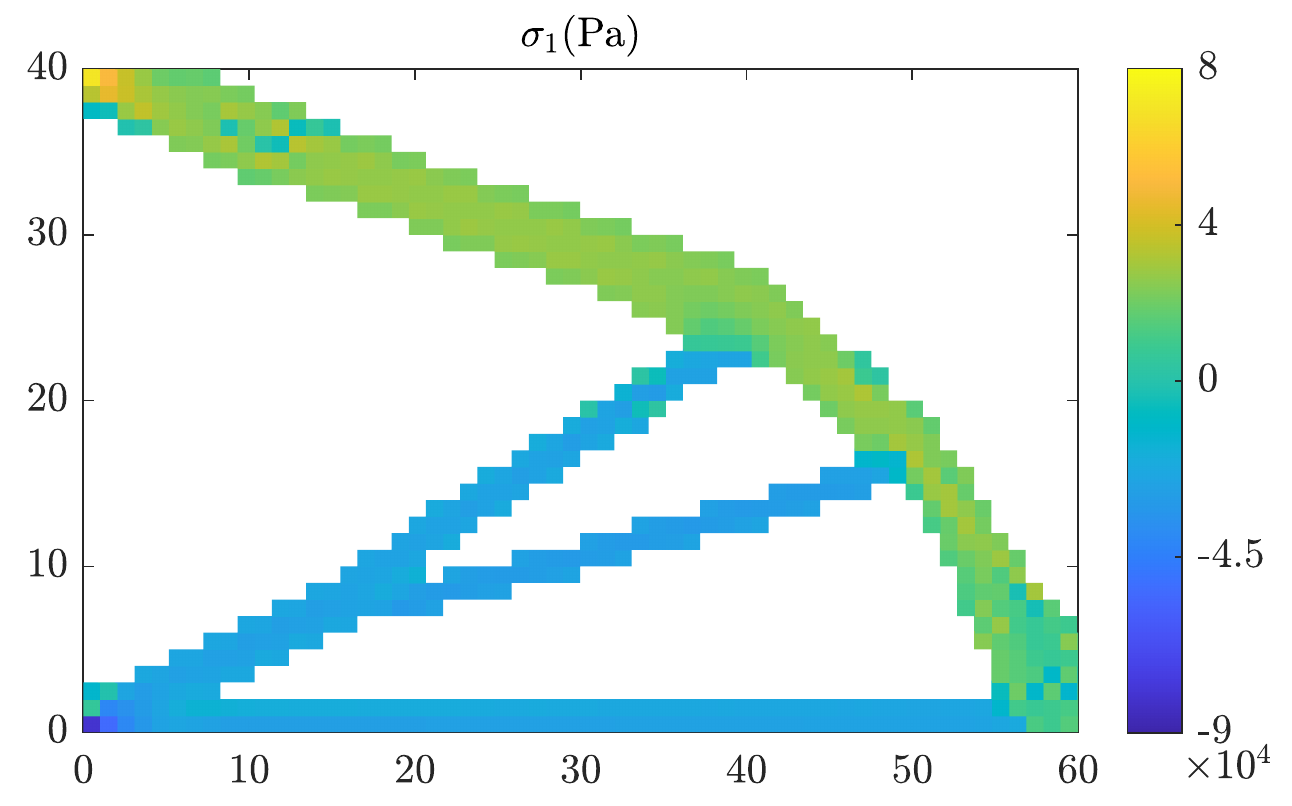}}
	\label{fig:8sx}
	\subfloat[$P=10$]{\includegraphics[width=0.99\columnwidth]{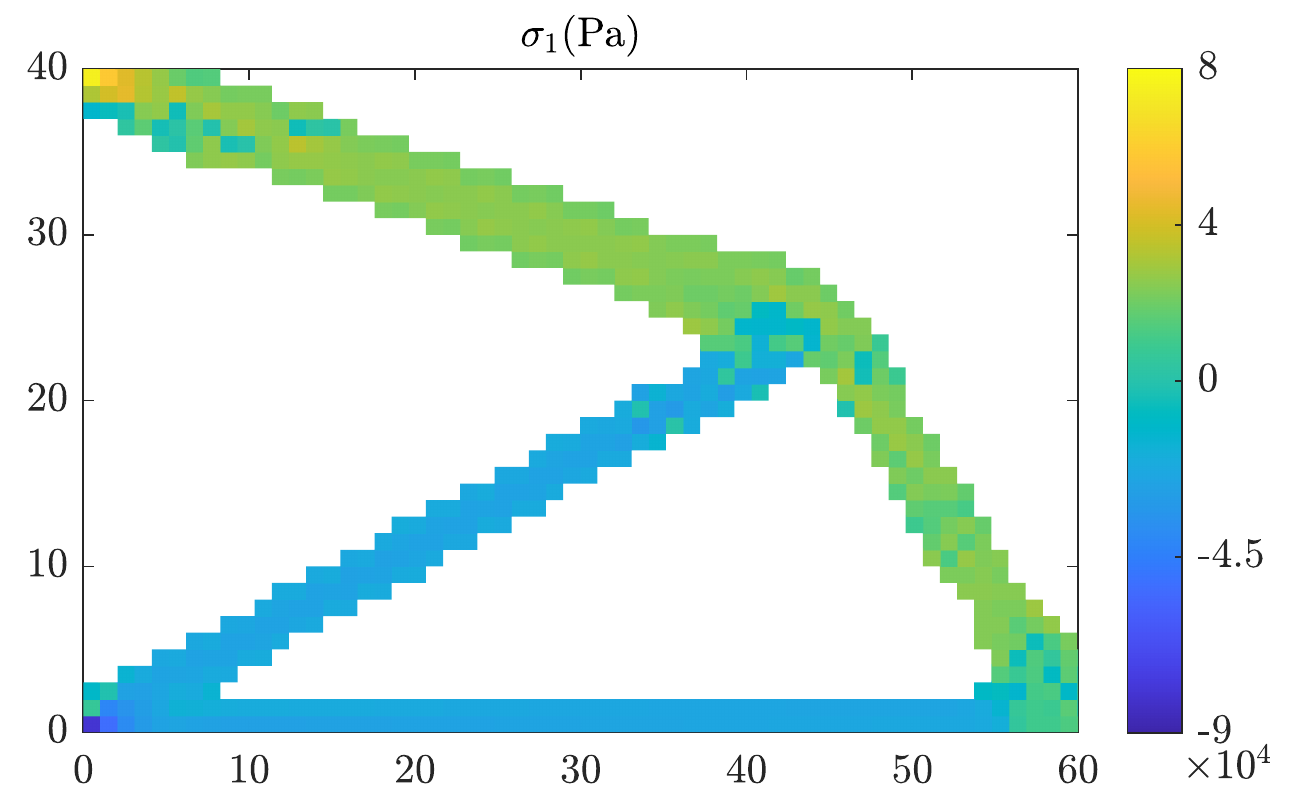}}
	\label{fig:10sx}
	\caption{Effect of $P$-value on stresses along the fibers (case study 4).}
	\label{fig:ch5p3sx}
\end{figure*}

\begin{figure*}
    \centering
	\subfloat[$P=4$]{\includegraphics[width=0.99\columnwidth]{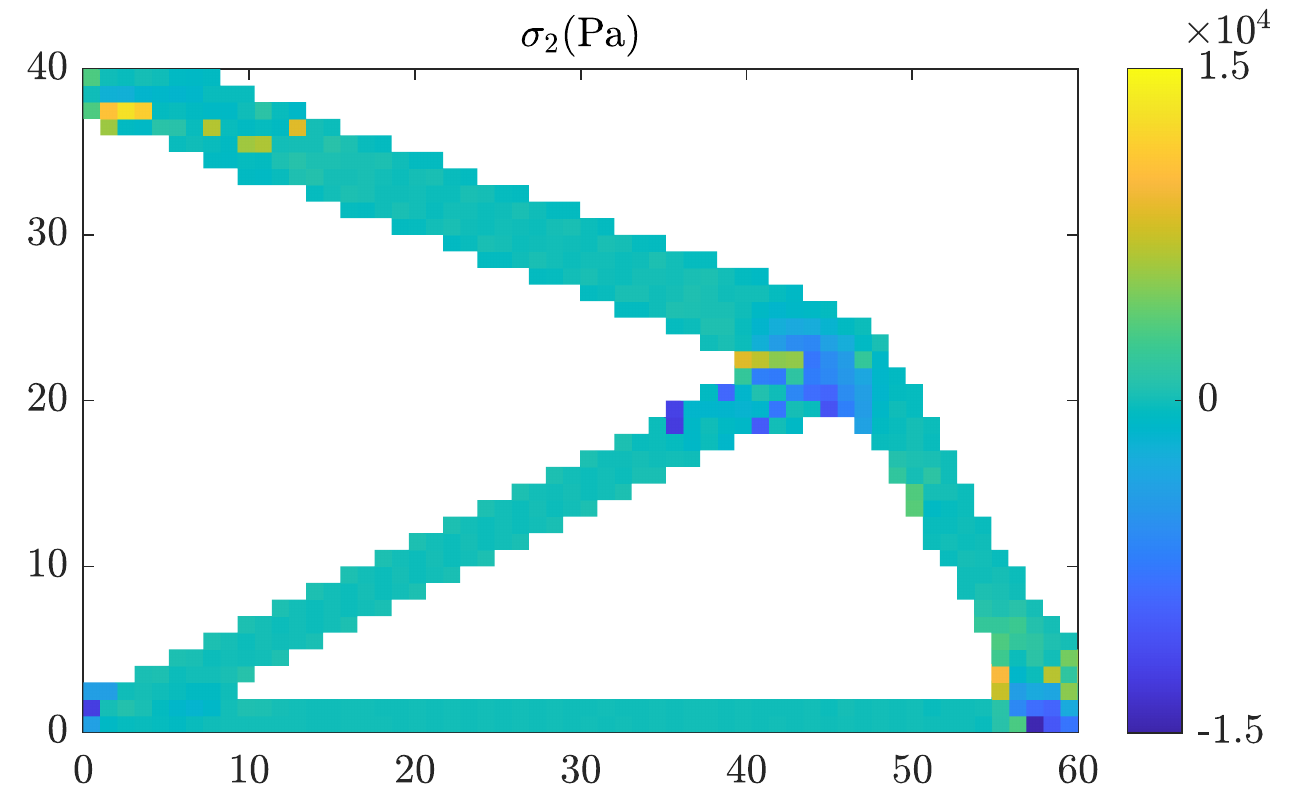}}
	\label{fig:4sy}
	\subfloat[$P=6$]{\includegraphics[width=0.99\columnwidth]{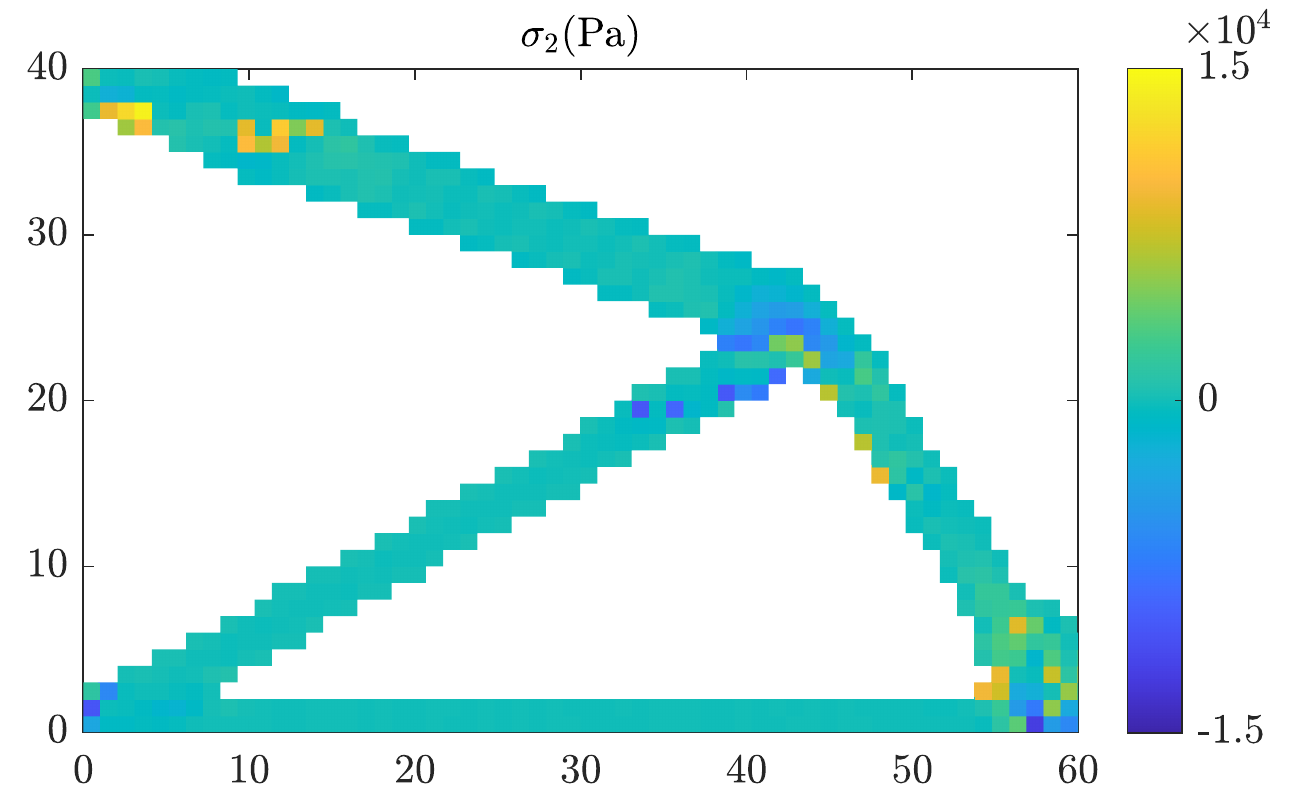}}
	\label{fig:6sy} \\
	\subfloat[$P=8$]{\includegraphics[width=0.99\columnwidth]{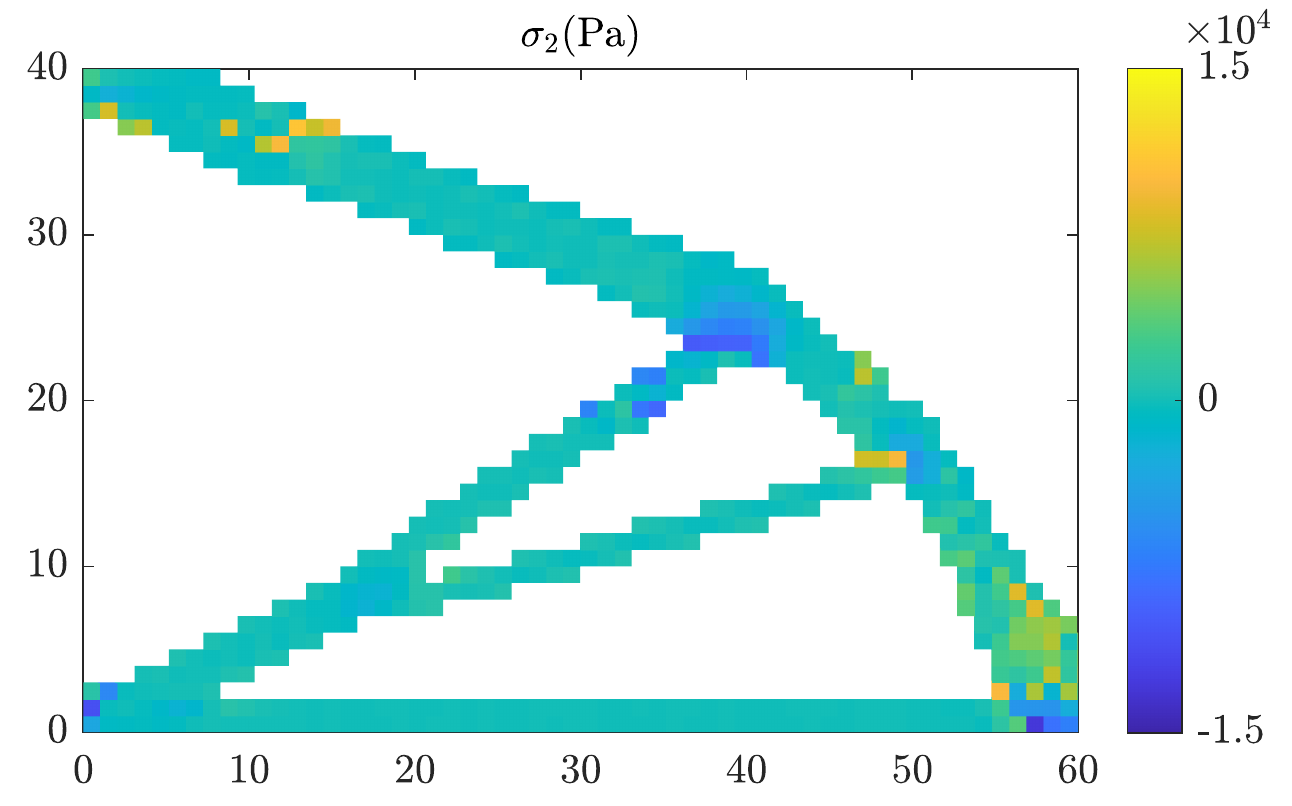}}
	\label{fig:8sy}
	\subfloat[$P=10$]{\includegraphics[width=0.99\columnwidth]{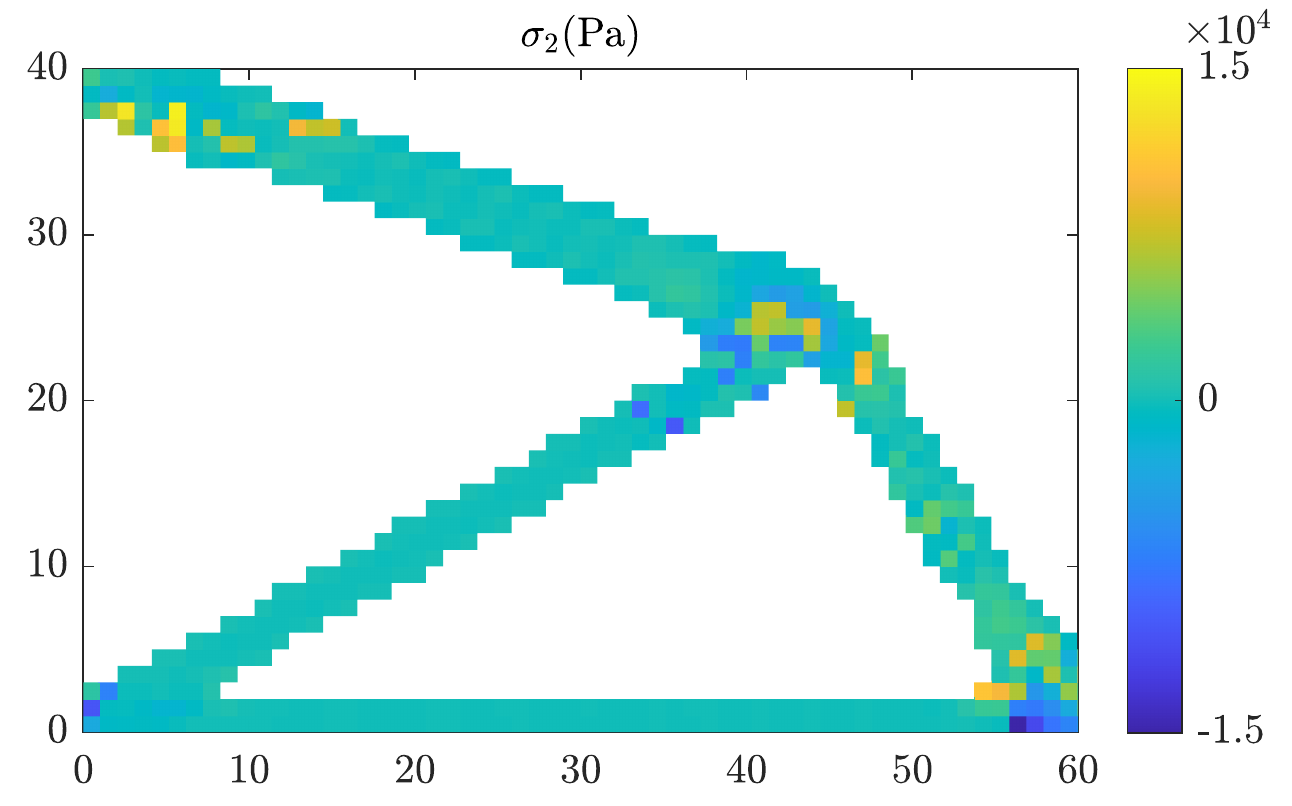}}
	\label{fig:10sy}
	\caption{Effect of $P$-value on stresses perpendicular to the fibers (case study 4).}
	\label{fig:ch5p3sy}
\end{figure*}

\section{Conclusions}
This paper proposes an efficient approach to stress constraining the topology and fiber orientation optimization problem of 3D printed structures. Constraining all the stress clusters is no longer necessary, since it causes numerical errors and inflates the compliance values. The numerical examples show that constraining only one or two clusters of stresses in each of the principal material coordinates is sufficient when using the stress level technique. This method produces structures that are innovative in the field of TO, since this is the first time that stress limitations were applied to orthotropic material. Using two clusters in each direction with a low number of stress evaluation points is better than using one cluster with a high number of stress evaluation points. The $P$-norm exponent value of 8 was the optimal value to use for this type of problems.
\vspace{1em}

\noindent \textbf{Acknowledgments}
We are grateful to Professor Svanberg for providing a version of the MMA code to be used for academic research purposes.
\vspace{1em}

\noindent \textbf{Replication of results} The authors have included all dimensions and material parameters needed to replicate the results.

\vspace{1em}
\noindent \textbf{\large Declarations}

\noindent \textbf{Conflict of interest} The authors declare they have no conflict of interest.
\vspace{1em}

\noindent \textbf{Funding} There is no funding source.
\vspace{1em}

\noindent \textbf{Ethical approval} This article does not contain any studies with human participants or animals performed by any of the authors.

\bibliographystyle{spbasic}
\pdfbookmark[0]{References}{References}
\bibliography{references}

\end{document}